\documentclass[usegraphicx, usenatbib]{mn2e}
\usepackage{times}

\title[The satellites of the Milky Way]{The satellites of the Milky Way - Insights from semi-analytic modelling in a $\Lambda$CDM cosmology} 
\author[E. Starkenburg et al.]{Else Starkenburg$^{1,2,3}$, Amina Helmi$^1$, Gabriella De Lucia$^4$, Yang-Shyang Li$^5$, 
\newauthor Julio F. Navarro$^2$, Andreea S. Font$^6$, Carlos S. Frenk$^7$, Volker Springel$^{8,9}$, 
\newauthor Carlos A. Vera-Ciro$^1$, Simon D. M. White$^{10}$ \\
$^1$ Kapteyn Astronomical Institute, University of Groningen, P.O. Box 800, 9700 AV Groningen, the Netherlands \\
$^2$ Dept. of Physics and Astronomy, University of Victoria, P.O. Box 3055, STN CSC, Victoria BC V8W 3P6, Canada \\
$^3$ CIfAR Junior Fellow and CITA National Fellow\\
$^4$ INAF - Astronomical Observatory of Trieste, via G. B. Tiepolo 11, I-34143 Trieste, Italy \\
$^5$ Kavli Institute for Astronomy and Astrophysics, Peking University, Beijing 100871, China \\
$^6$ Institute of Astronomy, University of Cambridge, Madingley Road, Cambridge, CB3 0HA \\
$^7$ Institute of Computational Cosmology, University of Durham, Science Laboratories, South Road, Durham DH13LE \\
$^8$ Heidelberg Institute for Theoretical Studies, Schloss-Wolfsbrunnenweg 35, 69118 Heidelberg, Germany \\
$^9$ Zentrum f\"ur Astronomie Universit\"at Heidelberg, Astronomisches Recheninstitut, M\"onchhofstr. 12-14, 69120 Heidelberg, Germany \\
$^{10}$ Max Planck Institut f\"ur Astrophysik Karl-Schwarzschild-Str. 1 85741 Garching, Germany}

\begin{document}

\maketitle

\begin{abstract}
We combine the six high-resolution Aquarius dark matter
simulations with a semi-analytic galaxy formation model to investigate
the properties of the satellites of Milky Way-like galaxies. We find good
correspondence with the observed luminosity function, luminosity-metallicity
relation and radial distribution of the Milky Way satellites. The star
formation histories of the dwarf galaxies in our model vary widely, in
accordance with what is seen observationally. Some systems are dominated by
old populations, whereas others are dominated by intermediate populations; star formation histories can either be
continuous or more bursty. Ram-pressure stripping of hot gas from the satellites leaves a clear imprint of the environment on the characteristics of a
dwarf galaxy. We find that the fraction of satellites dominated by old populations of stars matches observations well. However, the internal metallicity distributions of the model satellites appear to be narrower than observed. This may indicate limitations in our treatment of chemical enrichment, which is based on the instantaneous recycling approximation. We find a strong correlation between the number of satellites and the dark matter mass of the host halo. Our model works best if the dark matter halo of the Milky Way has a mass of $\sim8 \times 10^{11} \rm{M}_{\odot}$, in agreement with the lower estimates from observations, but about a factor of two lower than estimates based on the Local Group timing argument or abundance matching techniques. The galaxy that
resembles the Milky Way the most also has the best matching satellite
luminosity function, although it does not contain an object as bright as the SMC or LMC. Compared to other semi-analytic models and abundance matching
relations we find that central galaxies reside in less massive haloes, but the halo mass-stellar mass relation in our model is consistent both with hydrodynamical simulations and with recent observations. 
\end{abstract}

\begin{keywords}
galaxies: evolution -- galaxies: formation -- galaxies: dwarf -- galaxies: abundances -- galaxies: stellar content -- Galaxy: halo
\end{keywords}

\section{Introduction}

There is much to be learned about galaxy formation and evolution from our own `backyard', the Milky Way galaxy and its satellite system. Resolved stellar spectroscopy of the Milky Way stellar halo and the satellite galaxies provides `archaeological' evidence of the chemical enrichment of the interstellar medium back to the earliest times. The details with which the Milky Way and its satellites can be studied make them a useful testbed of the cosmological paradigm. 

The star formation history of the Milky Way satellites can be derived from the study of stellar populations identified in colour magnitude diagrams (CMD) \citep[see][and references therein]{tols09}. These studies have revealed large variations in the star formation histories of Local Group dwarf galaxies, even for those of similar stellar mass. These range from solely old- to predominantly intermediate-age to even significantly young stellar populations. Some star formation histories may be bursty, such as for the Carina dSph \citep{hurl98}. It is currently not completely understood what physical mechanisms are responsible for the exact star formation history and how the environment of the galaxy influences the star formation process.

With additional spectroscopic observations, several teams
 have investigated the
dynamical and chemical properties of both classical \citep[e.g.][]{tols06,batt06,koch07a,walk09,kirb10} and the recently
discovered ultra-faint dwarf spheroidal galaxies
\citep[e.g.,][]{kirb08,aden09,norr10}. The discovery of this new, very
faint class of satellites \citep{will05,zuck06a,zuck06b,belo07,wals07,irwi07} has revived the interest in the
so-called `missing satellites' problem \citep{klyp99,moor99}, which contrasts
the huge number of dark matter satellites predicted to orbit in Milky Way-sized
haloes with the relatively modest number of luminous satellites observed. 

In parallel to observational efforts, large cosmological N-body dark matter only
simulations, like the Aquarius Project \citep{spri08a}, Via Lactea II
\citep{diem08} and GHALO \citep{stad09}, have greatly improved mass resolution
and have now reached a regime in which the formation and evolution of
(satellite) galaxies can be studied in exquisite detail down to this
ultra-faint regime in a $\Lambda$CDM universe. 

Following the early  suggestion by \citet{efst92} and \citet{kauf93} that the reionization of the intergalactic medium at high redshift could suppress the formation of faint galaxies, \citet{bull00} were able to show, using dark matter halo merger trees and a 1-D gas simulation, that the effects of reionization could indeed help  to reconcile the distribution of  subhalo circular velocities expected in the CDM cosmology with inferences from satellite data. \citet{bens02b} then developed a detailed treatment of reionization and, using a semi-analytic model of galaxy formation, showed that the combined effects of reionization and supernova feedback could account for the observed luminosity function of satellites in the Local Group. They also predicted the existence of a large population of ultra-faint satellites. Several recent semi-analytical and hydrodynamical studies have made use of the new generation of N-body simulations \citep[e.g.,][]{muno09,okam09,okam10,bush10,coop10,macc10,li10,wade11,font11,sawa11a,guo11} to confirm the importance of reionization, and also feedback mechanisms, to suppress the formation of small galaxies within all haloes and reproduce the observed number of dwarf satellites down to the ultra-faint regime.

However, many issues remain. Relevant questions to be asked are for instance: `How many satellite galaxies are still
undiscovered in the Milky Way stellar halo?' \citep[e.g.][]{kopo08}, `What was their time of infall?' \citep[e.g.,][]{li10,roch11},
`What is the mass of the Milky Way dark matter halo?'
  \citep[e.g][]{batt05,sale07,smit07,liwh08,xue08,guo10}, `Are the luminosity
functions of satellites linked to the properties of their host in any way?'
\citep[e.g.][]{mcco06,mcco09,guoquan11,lare11,wang12a}.

In this work, we study the formation and evolution of dwarf galaxies in and around Milky Way-like galaxies using the N-body simulations of the Aquarius Project \citep{spri08a}. We combine these with semi-analytical modelling to study the physical processes associated with the baryonic components of the galaxies. We use the model described by \citet{li10}, which has been extended to include new prescriptions to follow the stellar stripping and tidal disruption of satellites. 

\citet{font11} also combined the Aquarius simulations with a semi-analytical code to study the properties of satellite galaxies, but their focus was in particular on a more sophisticated treatment of reionization, while our interest is mainly in the star formation histories of the satellites and of isolated dwarf galaxies. Because the two codes were developed independently, it is instructive to compare their results on general properties for the satellite population, such as luminosity function, metallicity distribution and radial profile. A similar semi-analytical model to that of \citet{font11} was used by \citet{coop10} to study the stripping of satellite galaxies and the formation of Galactic haloes in the Aquarius simulations. Additionally, \citet{guo11} have used an adapted version of the semi-analytical code of \citet{delu07} to study galaxies and satellites in the Millennium II simulation and show their results to be consistent with the satellite luminosity function over the (lower) resolution range in that simulation.

This paper is structured as follows. In Sections \ref{sec:Aq} and \ref{sec:SAM} we describe the Aquarius simulations and the particular semi-analytical model we use. Some additional prescriptions have been implemented to account for the tidal stripping and disruption of satellites, and these are described in detail in Appendix \ref{sec:strip}. In Section \ref{sec:genprop}, we investigate the properties of the modelled main galaxies as well as the luminosity function, the luminosity-metallicity relation, and the radial and spatial distributions of their satellites. Section \ref{sec:sfh} is devoted to a more in-depth analysis of the star formation histories of the modelled dwarf galaxies, both satellites and isolated galaxies, whereas in Section \ref{sec:findSclFnxCar} we investigate the closest model analogs to the dwarf spheroidal galaxies Sculptor, Carina and Fornax. In Section \ref{sec:disc} we discuss our findings and compare them with other semi-analytical and hydrodynamical work. We summarise our results in Section \ref{sec:Aqconc}.
 
\section{The model}

\subsection{The Aquarius simulations}\label{sec:Aq}
The six Milky Way-like haloes (Aq-A to Aq-F) of the Aquarius Project were selected from a lower resolution version of the Millennium-II Simulation \citep{boyl09}, a cosmological N-body simulation of a cubic region $125 h^{-1}$Mpc on a side with parameters $\Omega_{\rm{m}}=0.25$, $\Omega_{\Lambda}=0.75$, $\sigma_{\rm{8}}=0.9$, $n_{\rm{s}}=1$, $h=0.73$ and $H_{\rm{0}}=100h$ km s$^{-1}$Mpc$^{-1}$. We refer the reader to \citet{spri08a,spri08b} for further information. The parameters are the same as those of the Millennium Simulation \citep{spri05} and were based on the first-year results from the WMAP satellite. They are no longer consistent with the latest WMAP analysis \citep{koma11}, but we do not expect this to affect our results significantly \citep[see][for a comparison of first and third year parameters]{wang08}. The simulated Milky Way-like haloes have virial masses ($M_{\rm{200}}$, defined as the mass enclosed in a sphere with mean density 200 times the critical value) in the range 0.8--1.9$\times$$10^{12}$M$_{\odot}$, broadly consistent with the mass estimated for the Milky Way \citep[e.g.][]{batt05,smit07,liwh08,xue08,guo10}. One halo, Aq-A, was simulated at five different numerical resolution levels (summarised in Table \ref{Aqhalos}). We focus on the high-resolution level 2 (common to all six haloes) and use the lower resolution series Aq-5 through Aq-2 to test the numerical convergence of our model. 

Dark matter haloes are identified in the simulations using a friends-of-friends algorithm \citep{defw85} and the code {\sc subfind} \citep{spri01} which identifies self-bound structures within larger structures. Following previous work, we have only considered subhaloes that retain at least 20 particles. 

\begin{table}
\begin{tabular}{lcccc}
\hline
Name & $m_{\rm{p}}$ & $M_{\rm{200}}$ & $r_{\rm{200}}$ & Nr. snapshots\\
     & ($\rm{M}_{\odot}$) & ($\rm{M}_{\odot}$) & (kpc) & \\
\hline
\hline
Aq-A-5 & $3.14 \times 10^{6}$ & $1.85 \times 10^{12}$ & 246.37 & 128 \\
Aq-A-4 & $3.93 \times 10^{5}$ & $1.84 \times 10^{12}$ & 245.70 & 1024 \\
Aq-A-3 & $4.91 \times 10^{4}$ & $1.84 \times 10^{12}$ & 245.64 & 512 \\
\hline
Aq-A-2 & $1.37 \times 10^{4}$ & $1.84 \times 10^{12}$ & 245.88 & 1024\\
Aq-B-2 & $6.45 \times 10^{3}$ & $8.19 \times 10^{11}$ & 187.70 & 128\\
Aq-C-2 & $1.40 \times 10^{4}$ & $1.77 \times 10^{12}$ & 242.82 & 128\\
Aq-D-2 & $1.40 \times 10^{4}$ & $1.77 \times 10^{12}$ & 242.85 & 128\\
Aq-E-2 & $9.59 \times 10^{3}$ & $1.19 \times 10^{12}$ & 212.28 & 128\\
Aq-F-2 & $6.78 \times 10^{3}$ & $1.14 \times 10^{12}$ & 209.21 & 112 \\
\hline
\end{tabular}
\caption{Some basic parameters for the Aquarius haloes from \citet{spri08a} (see the original paper for more information). The columns correspond to the simulation name, the particle mass ($m_{\rm{p}}$), the virial mass of the halo ($M_{\rm{200}}$), the corresponding virial radius ($r_{\rm{200}}$) and the number of snapshots we use for each simulation.\label{Aqhalos}}
\end{table}

\subsection{The semi-analytical code}\label{sec:SAM} 
We use the Aquarius simulations as a backbone for modelling baryonic processes in galaxies. Subhalo catalogues are used to construct merger (history) trees for all self-bound haloes and subhaloes in the Aquarius simulations \citep{spri05,delu07} by determining one unique descendant for each (sub)halo. These merger trees are combined with semi-analytical modelling to study the galaxies that reside in such subhaloes. The semi-analytical modelling technique follows the relevant physical processes using simple but observationally and astrophysically motivated `prescriptions'. One advantage of this method is that it is applicable to large cosmological simulations and provides relatively fast predictions of galaxy properties. The method, however, does not follow explicitly the gas dynamics (as is done in hydrodynamical simulations), and does not usually provide spatially resolved information about the baryonic components.

The specific model we use in this work is the `ejection model' described in \citet{li10}, with new prescriptions to follow the stellar stripping and tidal disruption of dwarf galaxies as they become satellites. These new prescriptions are described in Appendix \ref{sec:strip}. The model builds upon the methodology introduced by \citet{kauf99}, \citet{spri01} and \citet{delu04} and subsequently updated by \citet{crot06} and \citet{delu07}. The model of \citet{delu07} has since been modified to follow more accurately processes on the scale of the Milky Way and its satellites by \citet{delu08,li09,li10}. In particular, both \citet{li10} and this paper use the `ejection' feedback scheme of \citet{delu04}, which is different from the (default) feedback scheme adopted in most previous work \citep[e.g.,][]{crot06,delu07,delu08}, a somewhat earlier reionization epoch and additionally suppress cooling in small haloes ($T_{\rm{vir}} < 10^{4}$ K). \citet{guo11} show that the model of \citet{delu07} significantly overpredicts the number of galaxies with stellar masses between $10^{7}$ and $10^{10} \rm{M}_{\odot}$. Our model is based on \citet{delu07}, but as outlined above several changes have been made since that paper. Most changes are focused on improving the modelling on low mass scales and are shown to mostly affect the dwarf galaxy scale and preserve the properties of galaxies with a Milky Way-like mass \citep[see Table 2. of][for a comparison]{li10}. However, \citet{delu07} show that the choice of feedback scheme will also affect the luminosity and evolutionary rate of the brightest cluster galaxies. In particular, the feedback model of \citet{delu04}, as used in this work, will result in a more prolonged star formation activity and a higher luminosity for these massive galaxies compared to the feedback model used in \citet{crot06}. A careful analysis of the combined impact of all the model changes on more massive galaxies will be part of future work. 

\begin{figure*}
\includegraphics[width=0.7\linewidth]{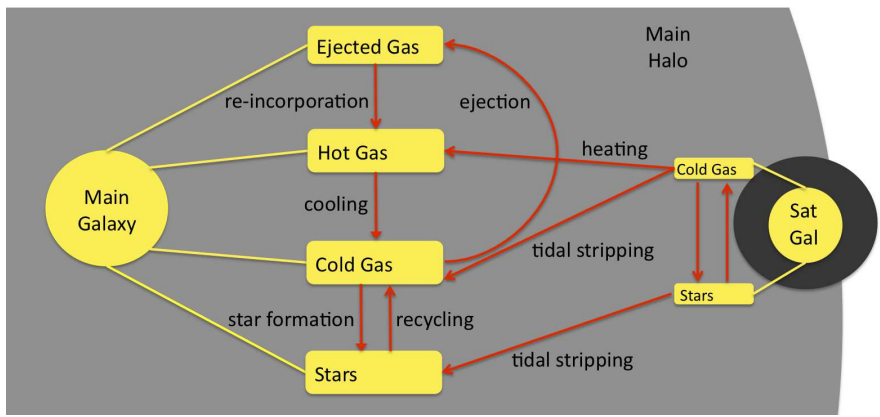}
\caption{A schematic diagram of the semi-analytical modelling scheme for the central galaxy within a dark matter halo and a satellite that has just been accreted. The yellow boxes linked to the galaxies represent all different `phases' of the baryons (these are all modelled analytically) and the red arrows represent all modelled physical prescriptions that affect them. \label{scheme}}
\end{figure*} 

A schematic diagram of the main processes modelled, is shown in Figure
\ref{scheme}. Below, we outline the main ingredients of the model, but
for a full description and the analytic expressions we refer the reader to
\citet{li10} and references therein.

\begin{itemize}
\item{\textit{Reionization} is modelled following \citet{gned00}, and \citet{crot06}. Reionization causes the baryonic content of a halo to decrease in haloes with mass comparable or smaller than a so-called filtering mass, which evolves with redshift. The reionization epoch is assumed to last from $z_{\rm{0}} = 15$ to $z_{\rm{r}} = 11.5$. Our reionization prescription results in a stronger effect than suggested for the global reionization in \citet{okam08}. However, \citet{font11} show using a detailed treatment of reionization for the Aquarius simulations that the proto-Galactic region is completely photo-ionized by $z=10$ due to the contribution of local sources. In the end, this results in a comparable reionization history on the scale of the Milky Way to that in our model \citep[see Appendix A of][and references therein for a complete discussion]{font11}.}
\item{\textit{Cooling} of the hot gas is dependent on its metallicity and temperature. Below $T_{\rm{vir}} = 10^{4}$ K (the atomic hydrogen cooling limit) cooling is forbidden since we are assuming that cooling via molecular hydrogen is prevented by photo-dissociation caused by UV radiation from the (first) stars in most cases \citep[in contrast to other semi-analytical work where star formation in haloes with $T_{\rm{vir}} < 10^{4}$ K is allowed through molecular hydrogen cooling; see for instance,][]{mada08, salv08}.}
\item{\textit{Star Formation} transforms cold gas, which is assumed to be in an exponential thin disc with properties given by the formalism of \citet{mo98}, into stars. Stars form in the gas of the disc that is above a critical density threshold. The star forming disc radius, r$_{\rm{disc}}$, is assumed to be three scale-lengths. Assuming that the disc has a flat rotation curve with a rotational velocity equal to the circular velocity of the halo \citep[V$_{\rm{200}}$, see also][]{kauf96} and the gas velocity dispersion is 6 km s$^{-1}$, the critical density threshold is described by Equation \ref{sflaw} \citep{kenn89}.
\begin{equation}
\frac{\Sigma_{\rm{crit}}}{\rm{M}_{\odot}\rm{pc}^{-2}} = 0.59\frac{V_{\rm{200}}}{\rm{km s}^{-1}} / \frac{r_{\rm{disc}}}{\rm{kpc}} \\
\label{sflaw}
\end{equation}
\ \\
The star formation rate is then proportional to the amount of gas in this state. Star formation can also happen in bursts during minor or major mergers, when (part of) the cold gas of the merging galaxies is turned into stars. A Chabrier initial mass function \citep{chab03} and instantaneous recycling approximation are assumed and accordingly 43\% of the mass in stars formed at each time step is (instantaneously) recycled back into the gas phase, representing supernovae Type II events.}
\item{\textit{Heating and ejection of gas} is due to supernova feedback processes. We use the feedback prescription from the ejection model described in \citet{delu04}. The mass of gas reheated by supernovae depends on the depth of the halo potential well (i.e., $\propto 1/V_{\rm{200}}^{2}$), which implies that smaller haloes (with shallow potential wells) are more sensitive to the effects of feedback. The material reheated is put in an ejected component that can be reincorporated into the hot gas at later times.}
\item{\textit{Metals} are created in star formation events and follow the flow of the mass between different components. We assume an instantaneous recycling approximation that is appropriate only for elements produced by Type II supernovae. }  
\item{\textit{If a galaxy becomes a satellite} (i.e., its halo becomes a subhalo) we assume that its hot gas and ejected gas components are stripped and transferred to the central galaxy. This process crudely models the physical process of ram-pressure stripping of hot gaseous components. In this paper we have also added two physical mechanisms that only operate on satellite galaxies due to interactions with the host halo, \textit{stellar stripping and tidal disruption}. The implementation of these physical processes and their results are described and shown in Appendix \ref{sec:strip}. We find that our implementation of stellar stripping does not have a significant effect on the luminosity functions, it affects very few galaxies. The tidal disruption prescription allows us to deal with galaxies which have `lost' their dark matter subhalo in the simulation (when it is stripped down to fewer than the {\sc subfind} resolution of 20 particles) and decide whether the galaxies themselves should survive the tides of the main galaxy. We refer to these galaxies as `orphans', in contrast to satellites which still live in dark matter subhaloes of more than 20 particles. Our implementation of tidal disruption has a significant influence on the shape of the satellite luminosity function, as we will show in Section \ref{sec:lumfunc}. }
\end{itemize}

\section{Comparison to the Milky Way and its satellites: General Properties}\label{sec:genprop}

\subsection{Milky Way properties}\label{sec:MW}

\begin{figure}
\includegraphics[width=\linewidth]{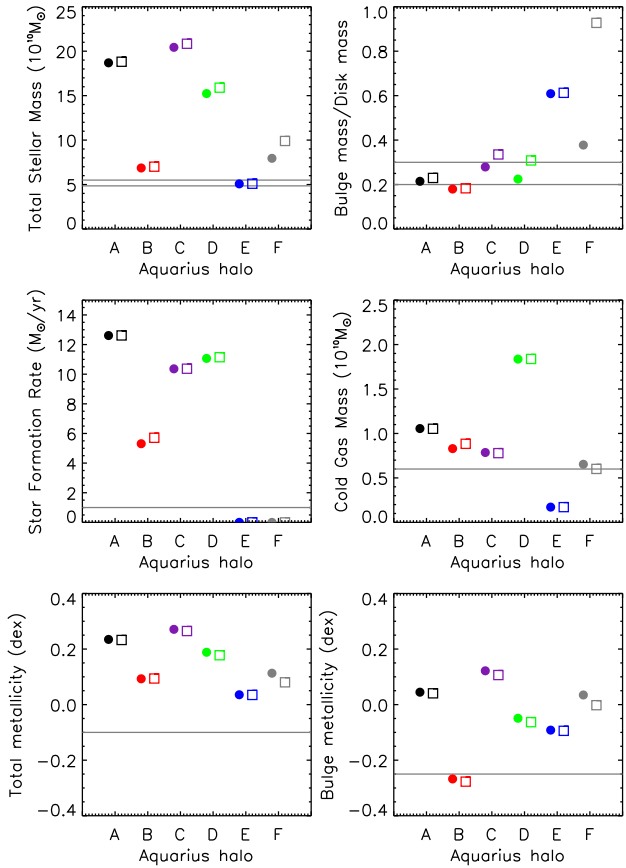}
\caption{Properties of the central galaxies of the haloes Aq-A to F both for the `satellite-model ejection' from \citet{li10} (filled circles) and our fiducial model (i.e., extended to include stellar stripping and tidal disruption, open squares). The solid lines show observed Milky Way values (see text for references). The properties of the Milky Way bulge are compared to the spheroid component in our models. \label{mainhalos}}
\end{figure} 

In Figure \ref{mainhalos} we show the various properties of the central (Milky Way-like) galaxies residing in the main Aquarius haloes compared to the values observed for the Milky Way. We show the results of our basic model \citep[][filled circles]{li10} and the same model with our additional prescriptions for stellar stripping and tidal disruption (see Appendix \ref{sec:strip}, open squares). Our treatment of satellite stripping and disruption changes only slightly the properties of the main galaxies. The scatter from main galaxy to main galaxy in the different Aquarius haloes is clearly visible.

Aquarius galaxies B and E have very similar stellar masses to the Milky Way, estimated to be in the range of $4.9 - 5.5 \times 10^{10}$ M$_{\odot}$ \citep{flyn06}, one of the most robust and important constraints for which the comparison can be made. 

In our models the spheroid refers to both the bulge and stellar haloes of the galaxies. Since in the case of the Milky Way the stellar halo contains a very small mass (in comparison to the bulge), we compare the spheroidal component in the model with the bulge component of the Milky Way. As explained in more detail by \citet{delu08}, spheroid formation can occur in our model through both mergers and disc instabilities. The treatment of disc instabilities is one of the least constrained physical processes in the model. It is very sensitive to small variations of the other prescriptions, but nevertheless is a key channel for bulge formation in Milky Way size galaxies \citep[see][for a comprehensive discussion]{parr09,delu11}. The disc and bulge components of the Milky Way are estimated to have a mass ratio between 0.2 and 0.3 \citep{biss04} which implies a bulge mass of $0.8 - 1.3 \times 10^{10}$ M$_{\odot}$. This bulge-to-disc ratio is well reproduced by the Aquarius galaxies, except in the case of galaxies E and F, which have a very dominant bulge component. 

The slight increase of the total spheroid mass (and thus also of the bulge/disc ratio) in the models with stripping and tidal disruption is expected since the stars stripped from the satellites are added to this component. Additionally, these processes can affect the details of the disc instability phenomenon through an increase in the cold gas mass of the disc. However, we find that for all haloes the major source of the bulge mass increase is the addition of stellar mass from disrupted and stripped satellites. For instance, the relatively large difference found in Aq-F is almost completely due to one quite luminous satellite galaxy that is tidally disrupted \citep[see][which includes a movie of the evolution of this object]{cooper11}.

In all models, except Aq-E which shows almost no star
  formation at the present day, we find that the current star formation rate is above the corresponding value for the Milky Way. This is also true for the total amount of
  present-day cold gas, which is also above the Milky Way value for most
  models (with the exception of Aq-F) \citep[using a total HI+H$_{2}$ mass value for
  the Milky Way of $\sim 6 \times 10^{9}$ M$_{\odot}$, see][ and references
  therein]{blit97}. Among the models showing significant ongoing star
  formation, halo B and F give the closest match to the Milky Way. 

The two bottom panels of Figure \ref{mainhalos} show the average metallicities obtained for all the stars and just the spheroid component. The overall metallicities lie systematically above the Milky Way values, but despite the crude approximations made to follow the evolution of metals in our model, the mismatch is only on average $\sim$0.25 dex, or a factor $\sim$1.8 in total mass of metals. Galaxy B, E, and F show the best match to the measured [Fe/H] for the disc (0.1-0.2 dex above the measured value), and halo B a very close match to the bulge stars \citep{free02}.

Altogether, Figure \ref{mainhalos} suggests that the galaxy modelled within halo B is the closest analogue of the Milky Way galaxy.

\subsection{Satellite luminosity functions}\label{sec:lumfunc}

\begin{figure}
\includegraphics[width=\linewidth]{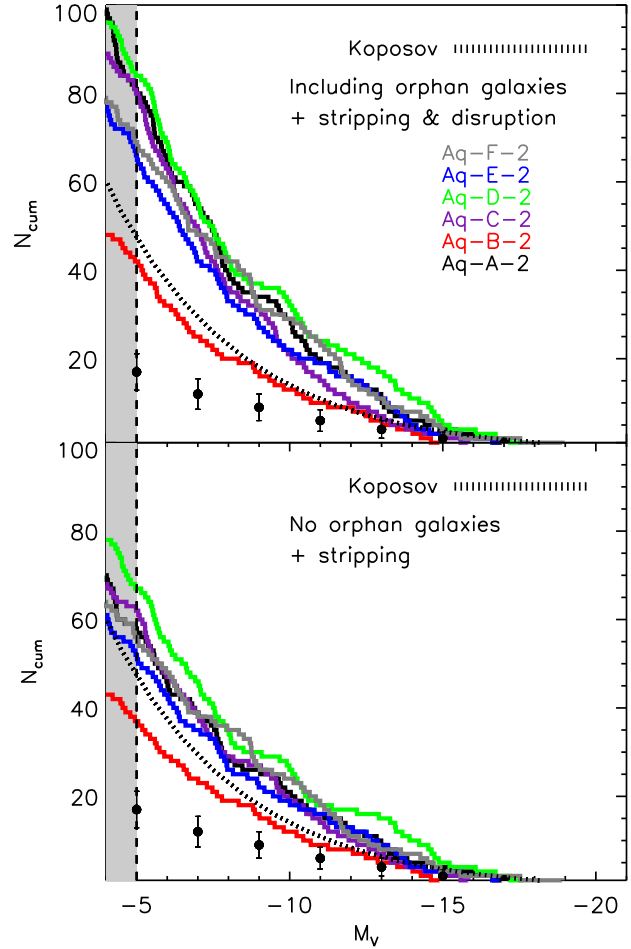}
\caption{Cumulative luminosity functions of all satellite galaxies within 280 kpc of the main galaxy for all different Aquarius haloes (solid lines), for the Milky Way satellites (black filled circles), with corresponding Poisson error bars, and for the Milky Way satellites as derived and corrected for incompleteness by \citet{kopo08} (thick dashed black line). In the top panel surviving orphan satellite galaxies are also included, in the bottom panel they are not shown.\label{lumfuncABCDEF}}
\end{figure}

Figure \ref{lumfuncABCDEF} shows the luminosity function of all satellites of the main Aquarius galaxies (solid lines of different colours). Black circles show the cumulative number of Milky Way satellites. Given that the Sloan Digital Sky Survey (SDSS) footprint covers roughly a quarter of the sky, we may expect many more galaxies hiding in the Milky Way that are either outside the SDSS footprint or too faint and/or too far away to be identified in SDSS data. The \citet{kopo08} (thick dashed black line) relation takes these uncertainties into account and attempts to correct for them. However, it should be noted that for the brightest end (M$_V < -11$) \citet{kopo08} use an average luminosity function of the Milky Way and M31 for their fit. Since M31 has more bright satellites than the Milky Way, the relation overpredicts satellites in this regime.

In the top panel of Figure \ref{lumfuncABCDEF} both the satellites still embedded within a dark matter subhalo and the orphan satellites (for which their dark matter subhalo has been stripped below the resolution limit of the simulation) which survive according to our tidal disruption prescription (see Appendix \ref{sec:strip}) are plotted. In the bottom panel these orphan satellites are not shown. Both the data and the model counts of satellite galaxies are restricted to a distance of 280 kpc from the centre of the main galaxy, as assumed by \citet{kopo08}. This radius is comparable to, but generally a bit larger than, the virial radii ($r_{\rm{200}}$) of the main Aquarius haloes (see Table \ref{Aqhalos}).

The shape of the luminosity function roughly agrees with the Milky Way data down to M$_V=-5$ where resolution effects are starting to play a role (see Appendix \ref{sec:numres} for the discussion of this limit, the vertical dashed line in Figure \ref{lumfuncABCDEF})
for all Aquarius haloes. Halo B shows clearly the best quantitative
correspondence with the luminosity function derived by
\citet{kopo08} in our fiducial model (top panel). It is interesting that Aquarius B
has a similar satellite luminosity function as well as the central galaxy that most closely resembles the Milky Way galaxy, as shown in the previous subsection. However, one should bear in mind that the number of satellites formed around any halo is very sensitive to the choice of feedback scheme and reionization physics \citep[e.g.,][]{guo11,font11}. 

Most Aquarius haloes do not contain satellites as bright as the LMC. This has been investigated by \citet{boyl10} and \citet{bush10}, who estimate from dark matter simulations that the probability of finding both an LMC and SMC around a Milky Way-sized halo is $\sim10$ per cent (up to 25 per cent with a dependence on the exact Milky Way dark matter halo mass and environment). \citet{liu11} find from an analysis of Milky-Way-like hosts in the SDSS DR7 catalogue that only 3.5 per cent of them have two such bright satellites within 150 kpc of their host. A further warning against making too strong a statement about the bright end of the luminosity function comes from \citet{guoquan11}, who find that isolated host galaxies of luminosity comparable to the Milky Way and to M31 contain $\sim$2 times fewer satellites brighter than $M_{\rm{V}}=-$14 \citep[see also][]{lare11}. In our models, 4 out of 6 haloes have one satellite as luminous as or slightly more luminous than the SMC (the exceptions are haloes B and C). Aq-F hosts a satellite galaxy slightly brighter than the LMC and one of luminosity similar to the SMC. Halo D hosts three very luminous satellites, but all fainter than the LMC.

\begin{figure}
\includegraphics[width=\linewidth]{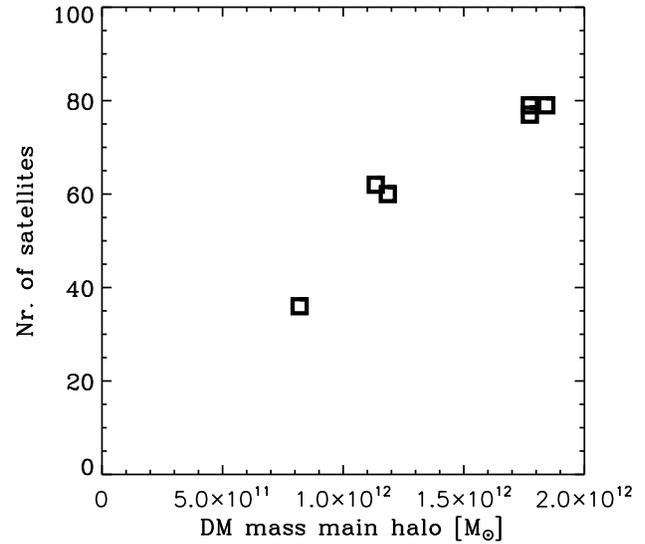}
\caption{The virial masses of the six different Aquarius dark matter haloes versus the number of satellite galaxies of the main galaxy in our model. Open black squares show the number of satellites brighter than M$_{\rm{V}}$=-5 within the corresponding virial radius.\label{massnum}}
\end{figure}

Although all Aquarius haloes have masses consistent with that of the Milky Way halo, they still span a factor 2.25 in mass. Aquarius B is the least massive with $M_{\rm{200}}= 8 \times 10^{11}$ M$_{\odot}$. Figure \ref{lumfuncABCDEF} shows that this range of masses is reflected in the number of satellites. This was noted also by \citet{macc10}, who remark that the trend between halo mass and satellite luminosity function does not depend on the particular semi-analytical model used. For our semi-analytical model this correlation is shown in Figure \ref{massnum}, where the number of satellites brighter than M$_{\rm{V}}$=-5  within the virial radius is plotted against the virial mass of the host halo. The trend is very clear and the scatter is small, although based on just a few points. We plan to investigate these results further using the much larger volume of the Millennium II simulation, where over 7000 Milky Way-sized haloes are found \citep{boyl10}.

Since our model for halo B best resembles both the properties of the Milky Way and the luminosity function of satellite galaxies, this would favour a dark matter mass estimate close to 8 $\times 10^{11}$ M$_{\odot}$ for the Milky Way galaxy in agreement with the work of \citet{batt05,smit07,sale07} and \citet{xue08}, but a factor two less massive than the best estimate of \citet{liwh08} and \citet{guo10}. In terms of formation history, halo B forms relatively late, its mass accretion is slower than that of other Aquarius haloes, in particular at $z>2$ \citep{boyl10}. In other properties, like spin or concentration, halo B is not special \citep{boyl10}.

\subsection{Luminosity-metallicity relation}\label{sec:lum-met}

\begin{figure}
\includegraphics[width=\linewidth]{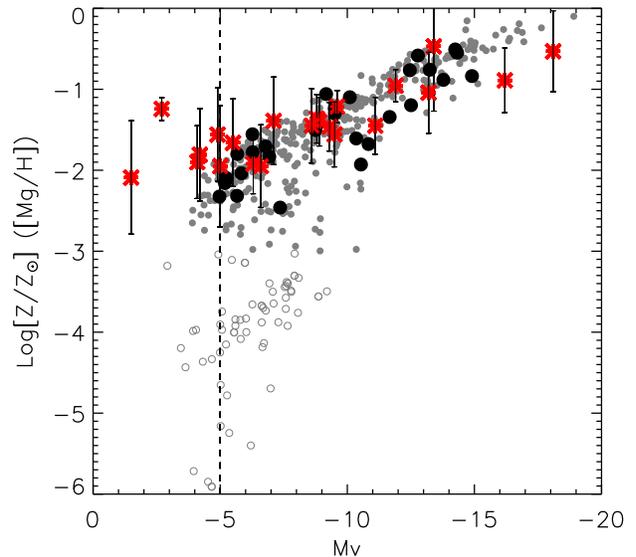}
\caption{Luminosity and metallicity for the satellite galaxies (grey filled circles) and those present only in halo B (black filled circles). Grey open circles represent satellite metallicities we do not trust, mainly due to incomplete modelling of first star physics. Overplotted as red asterisks are the average values for the Milky Way satellites, corrected to approximate a mass-weighted average of [Mg/H] for a better comparison to the models (see text for details and references). The error bars indicate the metallicity scatter found inside the galaxies. \label{lummet}}
\end{figure}

The Milky Way satellites \citep[including the ultra-faint dwarf spheroidals;][]{kirb08} show a strong correlation between luminosity and metallicity. It is not straightforward to compare the observed abundances to the metallicities in our model, since the model adopts an instantaneous recycling approximation that is valid for the majority of alpha elements formed like O and Mg, but not for Fe (mainly produced by Type Ia SN) which is the most commonly measured element in stellar spectra. 

For our comparison we therefore use Mg. Correcting the observed average [Fe/H] to [Mg/H] requires knowledge of [Mg/Fe]. Observations of red giant branch stars in dwarf galaxies and the Milky Way stellar halo show clear trends of [Mg/Fe] with [Fe/H], which vary slightly from satellite to satellite and are distinct from those in the Milky Way stellar halo, especially at [Fe/H]$>-1.5$. Nonetheless, we adopt the following function based on observed satellite data compiled by \citet{tols09} for [Mg/Fe] with [Fe/H]:

\begin{eqnarray}
\textnormal{[Mg/Fe]}=+0.4 \textnormal{\ \ \ \ \ \ \ \ \ \ \ \ \ \ \ \ \ \ \ \ \ for \ \ \ \ \ \ \ \ } \textnormal{[Fe/H]} < -2 \nonumber \\
\textnormal{[Mg/Fe]}=-0.4\textnormal{[Fe/H]}-0.4 \textnormal{\ \ \ \ \ for \ } -2 < \textnormal{[Fe/H]} < 0 \\
\textnormal{[Mg/Fe]}=-0.4 \textnormal{\ \ \ \ \ \ \ \ \ \ \ \ \ \ \ \ \ \ \ \ \ for \ \ \ \ \ \ \ \ \ \ } \textnormal{[Fe/H]} > 0\nonumber
\end{eqnarray}

  The metallicity given by our model is mass-weighted, i.e. it is the logarithm of the ratio of mass in metals over the total mass in stars. The average observed metallicity to compare with should therefore also be obtained by taking the logarithm of an average over the ratio of metals to hydrogen, which will be different from the average of [Fe/H]. We have tested the offset between the two estimates of the mean metallicity on the data set of the Dwarf Abundances and Radial velocity Team (DART) \citep{tols04}, which contains (Ca II triplet derived) metallicities for the classical dwarf spheroidals Fornax, Sculptor, Carina and Sextans. We found that the average [Fe/H] is on average 0.23 dex lower than the logarithm of the average over the ratio of Fe to H, although ranging from 0.15 to 0.4 dex. In Figure \ref{lummet} we therefore show [Fe/H] corrected by 0.23 dex and subsequently transformed to [Mg/H] values. For Fornax, Sculptor, Sextans and Carina we do not use the 0.23 correction, but use the averages directly from the DART data (see \citet{tols06}, \citet{batt08} and \citet{star10} for a description of the data sets and methods) by taking the logarithm of the average ratio of Fe to H for all stars.

Figure \ref{lummet} shows a comparison for those Milky Way satellites which have an average [Fe/H] available from the literature \citep[see, for instance, the compilation in][]{ysthesis}. The mean iron abundances and their dispersions (error bars) in the plot are taken from \citet{west97} for the LMC and SMC, \citet{cole01} for Sagittarius, the DART survey \citep[][and references therein]{helm06,star10} for Fornax, Sculptor, Carina and Sextans, \citet{harb01} for Draco and Ursa Minor, \citet{koch07a} for Leo I and \citet{koch07b} for Leo II, \citet{kirb08} for most of the ultra-faints; B\"ootes I and Segue I are from \citet{norr10} and B\"ootes II from \citet{koch09}. 

The filled and empty grey circles in Figure \ref{lummet} show the average metallicity for the model satellite galaxies within 280 kpc of their hosts. The larger black filled circles represent the satellites in Aq-B, to highlight the number of satellites and dispersion in metallicity found in that simulation (which most resembles the Milky Way). 

All model galaxies with an average metallicity below [Fe/H] =$-3$ are shown as open grey circles. All of them have experienced very few star formation events (typically less than 4). Because of their few star formation episodes these satellites will enrich very little and not compensate for the first generation of stars formed with no metals at all, in contrast to higher mass objects which can sustain star formation for more extended periods. Their very low metallicities could therefore result mainly from our neglect of any kind of pre-enrichment, which could be driven by a top-heavy IMF for the first stars. These events are likely to enrich the galaxy, or even the intergalactic medium to a metallicity floor of [Fe/H] $\sim -$3 \citep[e.g.][]{salv08}.

The metallicities for SMC and LMC-like satellites are overpredicted, although they are found to be consistent within the range of the measured scatter inside these galaxies. A more thorough modeling of the chemical processes will have to be conducted however, before enabling any conclusions on the possibility that our model retains too much of its produced metals for the more massive galaxies. The overall slope and normalisation of the metallicity-luminosity relation are reproduced well in our models when considering those galaxies which have sufficient star formation events to enrich above [Fe/H]=-3, as shown in the grey filled circles. 

\subsection{Radial and spatial distribution}

\begin{figure}
\includegraphics[width=\linewidth]{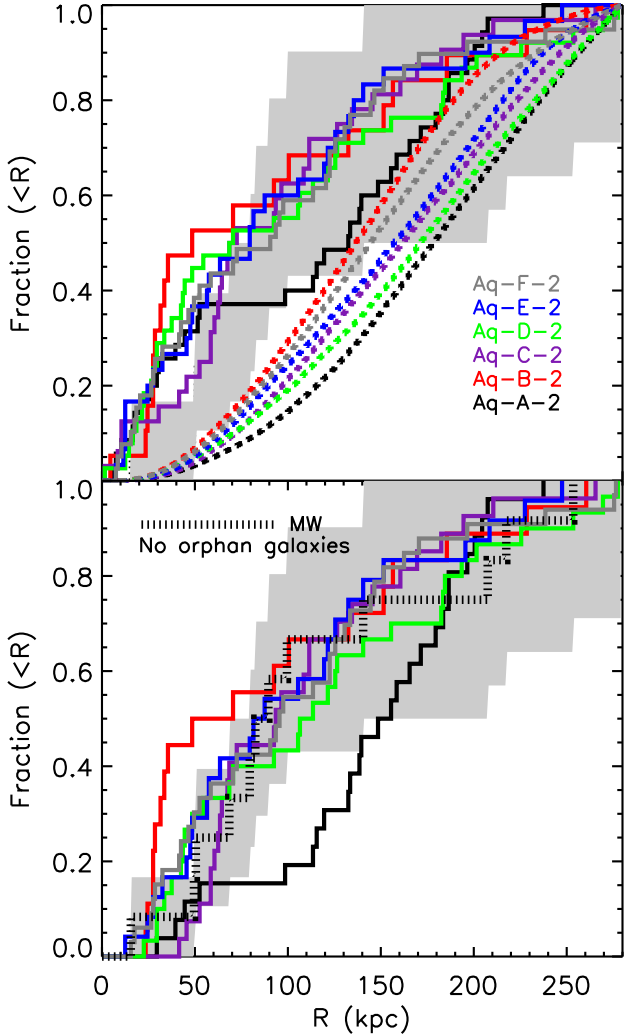}
\caption{Radial distribution of the bright model satellites (M$_{\rm{V}} < -8.5$, solid lines). The top panel shows all bright satellites (solid lines) as well as the total population of all subhaloes (dashed lines). The grey area indicates the distribution of Milky Way satellites including a Poissonian error bar in both panels. In the bottom panel additionally the radial distribution of classical Milky Way satellites is overplotted (dotted black line) and orphan satellites are excluded from the solid lines. \label{raddistr}}
\end{figure}

In Figure \ref{raddistr} we show the radial distributions of all satellites brighter than M$_{\rm{V}}$=$-8.5$ within the Milky Way and in each of the Aquarius haloes. For the (surviving) orphan satellites included in the top panel, the position is that of the most bound particle within the host subhalo before disruption. For the Milky Way satellites, the distances are taken from \citet{mate98}, except for Canes Venatici I \citep{mart08}.  We have assumed a distance from the Sun to the Galactic centre of 8.5 kpc. The uncertainty in the radial distribution due to Poisson noise is indicated by the grey area. 

The distribution of satellites is broadly consistent with that observed in the Milky Way for most Aquarius haloes, except for the inner regions. If orphan satellites are removed, the profiles are in general slightly less centrally concentrated, since the orphan galaxies are mainly found in the inner regions. The bottom panel of Figure \ref{raddistr} shows that the radial distributions of Aquarius C, D, E, and F match that of the Milky Way if orphan galaxies are excluded from the analysis. Such an exclusion could be justified, since orphan galaxies are almost certainly affected by tides and could have very low surface brightness, which would hinder their detectability. 

Note that the radial distribution of luminous satellites is different from that of the total population of dark matter substructures, which has a less centrally concentrated profile as shown as dashed lines in the top panel of Figure \ref{raddistr} \citep[see also][]{gao04, spri08a}.

\begin{figure}
\includegraphics[width=\linewidth]{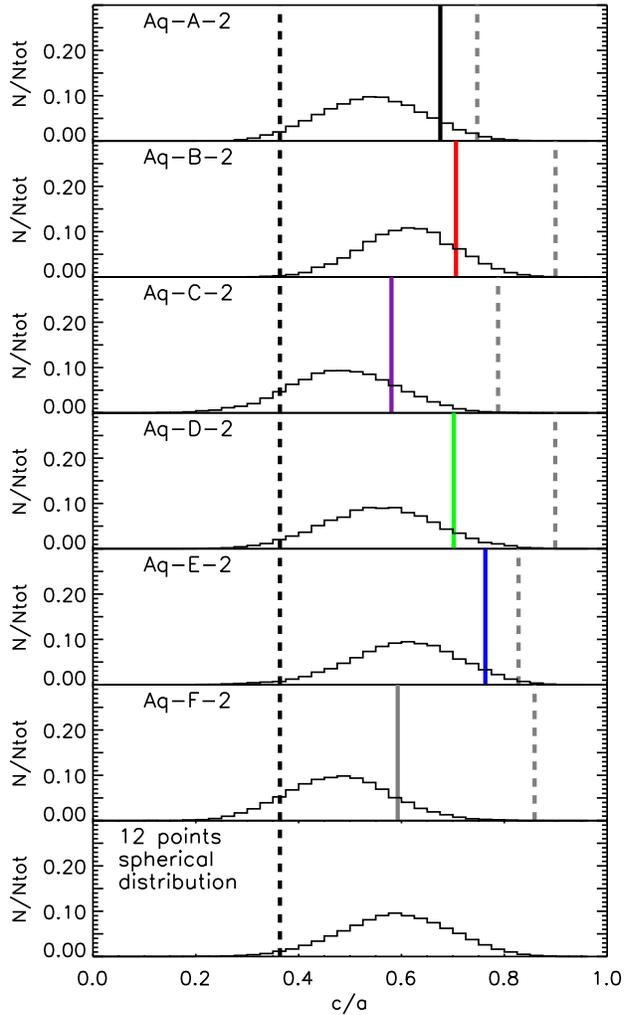}
\caption{Flattening (c/a) of the twelve classical satellites (M$_{\rm{V}} <$ -8.5) in the Milky Way (black dashed vertical line) and the distribution of c/a values that can be reached with random selections of twelve M$_{\rm{V}} <$ -8.5 model satellites for the Aquarius haloes A-F. For each Aquarius halo also the flattening of \textit{all} satellites with M$_{\rm{V}} <$ -8.5 is overplotted as a full coloured line. We also show the flattening of all subhaloes within 280 kpc of each main halo as a dashed grey line. The distribution of c/a obtained by taking twelve random points from a spherical distribution is shown in the bottom panel for comparison. \label{spatdistr}}
\end{figure}

One interesting property of the Milky Way satellites is that they seem to lie close to a plane, rather than being distributed isotropically on the sky \citep[e.g.][]{kunk76,lynd76,maje94,lynd95,hart00,palm02,krou05,metz07}. In Figure \ref{spatdistr} we investigate the anisotropy of the spatial distribution of satellites with  M$_{\rm{V}}<-8.5$ in each of the Aquarius haloes. We calculate the flattening using the normalised inertia tensor. The short-to-long axis ratio (c/a) is computed from the eigenvalues of the diagonalized inertia tensor. In each panel the flattening of the Milky Way satellite system is indicated by a dashed black vertical line, the flattening of the total system of subhaloes in each Aquarius halo as a dashed grey line, whereas that of all bright satellites is shown as a full coloured line. 

All Aquarius haloes show a fairly spherical distribution of their total system of (dark) subhaloes within 280 kpc, although some variations can be seen from halo to halo. The spatial distribution of their bright satellites is in all Aquarius haloes less flattened than the Milky Way satellite system. However, all host a larger number of bright satellites as well, as shown in Figure \ref{lumfuncABCDEF}. To investigate the effect of the number of satellites on this comparison, we have overplotted in all panels in Figure \ref{spatdistr} the distribution of c/a when instead of all bright satellites a random subsample of 12 satellites is taken, equal to the number of classical satellites in the Milky Way. The restriction to a smaller number of satellites greatly enhances the chances of selecting a more flattened distribution, as can also be concluded from the bottom panel where 12 points are selected randomly distributed on a sphere. From a purely spherical distribution one expects a flattening comparable to that seen in the Milky Way in $\sim$1 per cent of the cases if 12 points are drawn randomly. 

\begin{figure}
\includegraphics[width=\linewidth]{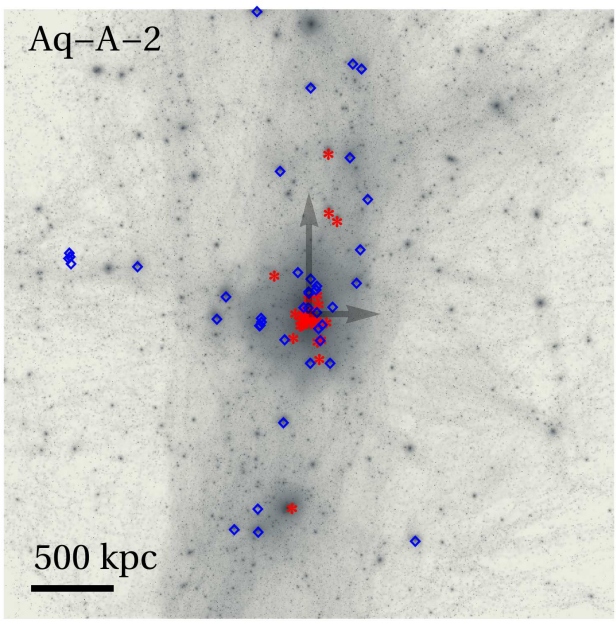}
\caption{The spatial distribution of galaxies of $-19<$M$_{\rm{V}}<-8.5$ in Aquarius A. Symbols and colours denote the age of the dominant stellar population: old (red asterisks), or intermediate (blue diamonds). The grey map shows the underlying density of dark matter in the same simulation. The frame is rotated such that the major axis of the main halo is vertical and the minor axis horizontal. The grey errors show the relative sizes of the major and minor axis. \label{oldyoung_spat}}
\end{figure}

The chance of getting such a highly flattened distribution as seen in the Milky Way from 12 satellites within the Aquarius models is low, but can also not be completely ruled out. In particular, Aquarius B and E show distributions close to the spherical case when 12 satellites only are selected. It is extremely unlikely for Aquarius B to host a Milky Way-like flattened system (but notice that Aq-B has significantly fewer bright satellites in total than the other haloes). Aquarius A, C and F are on average more flattened. In some of these cases the flattening of the satellite system follows the shape of the present-day host dark matter halo and/or large scale structure, most notably in Aquarius A as illustrated in Figure \ref{oldyoung_spat}. Aquarius A is found to have a long, thin filament which is coherent in time, whereas in some of the other Aquarius haloes the filament is either less well-defined or broader (such that it encompasses the whole halo) or changes its orientation over time \citep[for a full discussion of the shapes of the Aquarius dark matter haloes and their filaments we refer the reader to][]{vera11}. 

Other studies have also indicated that a flattening similar to that of the Milky Way satellites can be reached in a $\Lambda$CDM cosmology, although it is not very common \citep[e.g.][]{libe05,kang05,zent05,li08,libe09,deas11}. 

\section{Star formation histories}\label{sec:sfh}

\subsection{A comparison to Local Group dwarf galaxies}

The resolution for which the stellar ages can be determined depends on the populations that can be used as tracers. Overall, the resolution decreases with increasing age and ranges from a few Myr at the youngest end (ages up to 1 Gyr), to several Gyrs for stars older than a few Gyrs. In this work we consider three different age bins that can be well separated in a CMD analysis: an old population ($>$ 10 Gyr), an intermediate population (1-10 Gyr) and a young population, $<$1 Gyr \citep[see also][]{tols09}. 

\begin{figure}
\includegraphics[width=\linewidth]{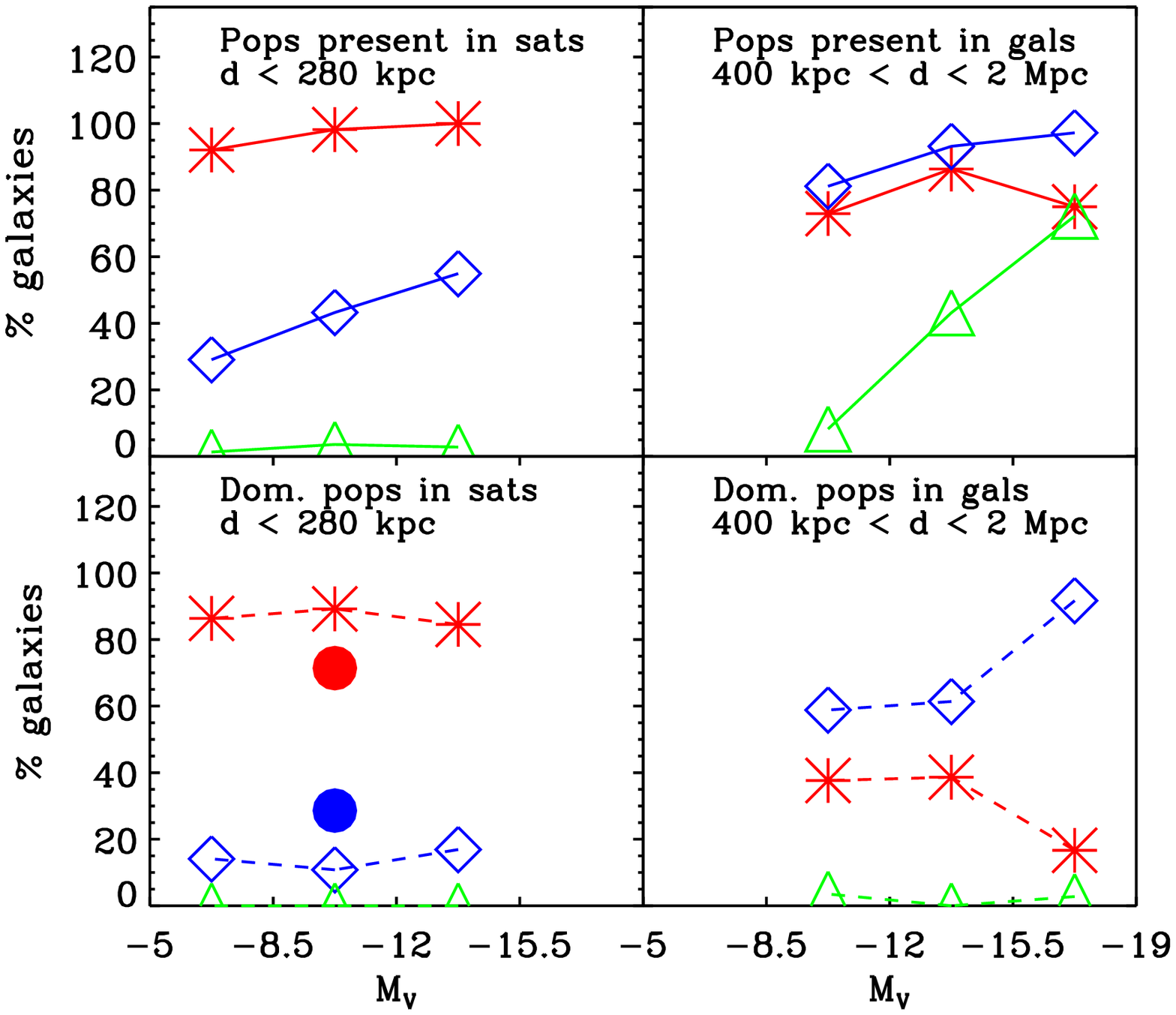}\\
\includegraphics[width=\linewidth]{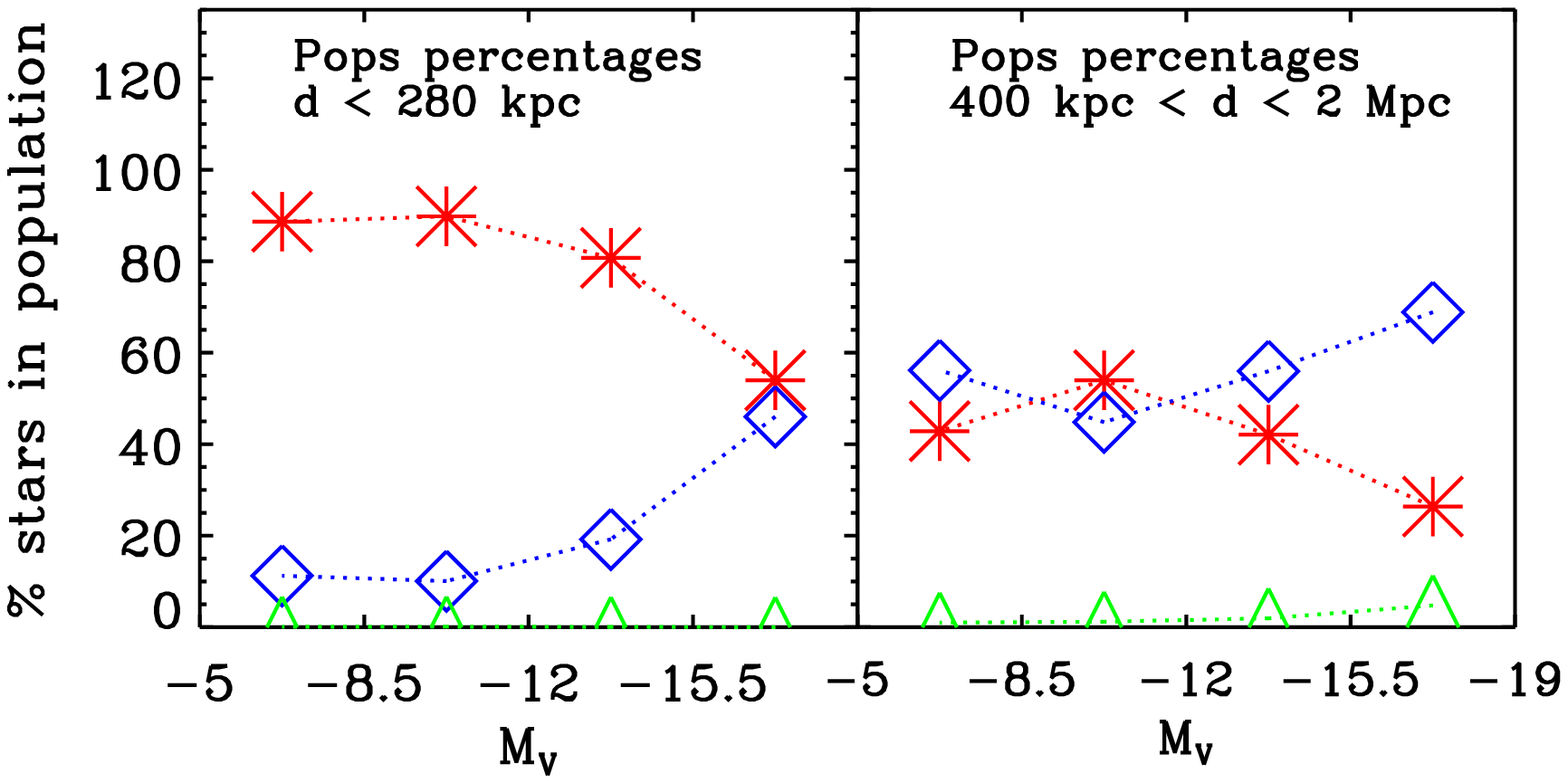}
\caption{The percentage of galaxies in each luminosity bin which have $>$1 per cent (top panels) old (red asterisks), intermediate (blue diamonds) and/or young (green triangles) populations. In the middle panels the lines with similar symbols indicate whether the old, intermediate or young populations are the dominant one in the galaxy (that is, more than 50 per cent of their stars originate from this epoch). In the bottom panels we show the fractions of the old, intermediate and young stars averaged over all galaxies in the luminosity bin. In all left panels the satellite galaxies, within 280 kpc of the main halo are displayed, in all right panels the isolated galaxies. Overplotted in the left middle panel are the percentages of these Milky Way satellites with $-12<$M$_{\rm{V}}<-8.5$ dominated by either old or intermediate population as filled red and blue circles respectively for a direct comparison. \label{oldyoung}}
\end{figure}

In the top left panel of Figure \ref{oldyoung} we show the percentage of satellite galaxies that contain observable populations (defined as $>$1 per cent of the total mass) in each of the age bins in the simulations Aq-A to F, where red asterisks correspond to the old, blue diamonds to the intermediate and green triangles to the young populations of stars. 

From this panel we see that $\sim$9 per cent of all modelled satellites do not have an old population, whereas all Milky Way satellites and all isolated dwarf galaxies observed with sufficiently deep colour-magnitude diagrams do contain old stars. We have checked explicitly that all of these subhaloes have appeared in the simulation well before a lookbacktime of 10 Gyrs, therefore the lack of old populations is probably related to the implementation of the semi-analytic prescriptions. There are at least two possible explanations for this difference associated to how we model star formation and cooling. 

Firstly, we do not allow cooling in dark matter haloes with T$_{\rm{vir}}<10^{4}$K. If, however, we relax this assumption and also allow cooling in such haloes with the same efficiency as a halo of T$_{\rm{vir}} = 10^{4}$K, we find that a larger percentage of all galaxies (95 per cent of all dwarf galaxies, and 97.5 per cent of those that are satellites presently) does form an old population. But as a consequence the faint end of the satellite luminosity function is also much enhanced (most notably at M$_{\rm{V}}>-5$). A proper implementation of the physical processes playing a role in the formation of the first stars would require many extra assumptions, for instance on the IMF and the interplay between H$_{2}$ cooling and H$_{2}$ dissociation in the host haloes which we prefer not to include in our models at this stage. 

Another limitation of our semi-analytical model is that we do not represent the stochasticity of star forming regions. In our modelling, for stars to form in a disc, the total (global) surface density of the disc has to be above the star forming density threshold.  However, in reality star formation processes can be much more local, i.e. one molecular cloud can have the required density while its surroundings might not. 

In the left middle panel we plot the percentage of galaxies for which a given population is dominant (i.e. more than 50 per cent of the stars belong to it) as a function of $M_{\rm{V}}$. Table \ref{tab:MWsats} summarizes the available data for the Milky Way satellites \citep[the first 5 columns are taken from Table 2 of][]{tols09}. A direct comparison can be made for galaxies in the luminosity bin $-12<$M$_{\rm{V}}<-8.5$. Most modeled satellites in this bin are dominated by an old population, while a significant minority ($\sim$20 per cent) is dominated by an intermediate population of stars. This is very comparable to the results for the satellite galaxies around the Milky Way, as can be directly seen from their percentages overplotted in the left middle panel of Figure \ref{oldyoung} as filled circles, where out of the seven galaxies in this luminosity bin only two are clearly dominated by an intermediate population. 

The left bottom panel of Figure \ref{oldyoung} shows the relative contributions of old, intermediate and young populations averaged for all galaxies in a particular luminosity bin.

While all left panels show the satellite galaxies of the main Aquarius galaxy, the right panels of Figure \ref{oldyoung} show the contributions of the various populations for dwarf galaxies outside the main halo, i.e., from 400 kpc to 2 Mpc. We focus here on slightly brighter dwarf galaxies, with $-19<M_{\rm{V}}<-8.5$, as these may be observable with current instrumentation. Most of these systems are the main galaxy within their dark matter halo, which are the only galaxies fed by cooling. The isolated dwarfs are mostly dominated by intermediate age instead of old populations. Also, a young population of stars is present much more frequently (compare the two top panels). Qualitatively, it is clear that our model produces an age-density relation in agreement with the observations of Local Group satellites. Gas-poor and older galaxies are found in overdense regions near bigger galaxies, while the star forming galaxies are found in isolation. This result is more clearly illustrated in Figure \ref{oldyoung_spat}, where we show the location of $-19 <$ M$_{\rm{V}} < -8.5$ galaxies overlaid on the dark matter distribution for the Aquarius A simulation. (We have checked that the results shown in Figure \ref{oldyoung_spat} are not influenced by projection effects.) The galaxies dominated by an old stellar population show a very different distribution as function of the density of their environment than the intermediate age galaxies. 

Some galaxies outside the main haloes are dominated by old populations. In all of these some stars are formed at intermediate ages as well, but there are fewer of these than  old stars. All show a bursty star formation at intermediate age. Some have never been satellites in their life, so this behaviour is entirely due to their own internal feedback processes. In the Local Group we also find a few examples of old and passively evolving small galaxies, like the Cetus and Tucana dwarf spheroidals, with no clear association to the Milky Way or M31, although a past association can not be ruled out. The existence, or non-existence, of truly isolated quenched dwarf galaxies in observational data will provide important constraints on the modelling of the star formation threshold for dwarf galaxies.   

A more quantitative comparison between isolated dwarfs in our models and in the nearby Universe is difficult. Many of the well studied galaxies outside the Milky Way are still associated with its nearest neighbour, Andromeda, or with the Local Group environment as a whole. Beyond the Local Group, at $\sim$ 1.3 Mpc, the  observational CMDs are much harder to interpret, since only the giant branch is bright enough to be resolved. We have made a tentative comparison with the ACS Nearby Galaxy Survey (ANGST) survey, a project measuring the star formation histories of galaxies outside the Local Group out to 4 Mpc in a systematic manner \citep{dalc09,weis11}. Only three galaxies have reliable measurements of the dominant population in their sample from 1.3 to 2 Mpc. Of these three galaxies, two are dominated by an intermediate population and just one by an old population \citep{weis11}. 

\begin{figure}
\includegraphics[width=\linewidth]{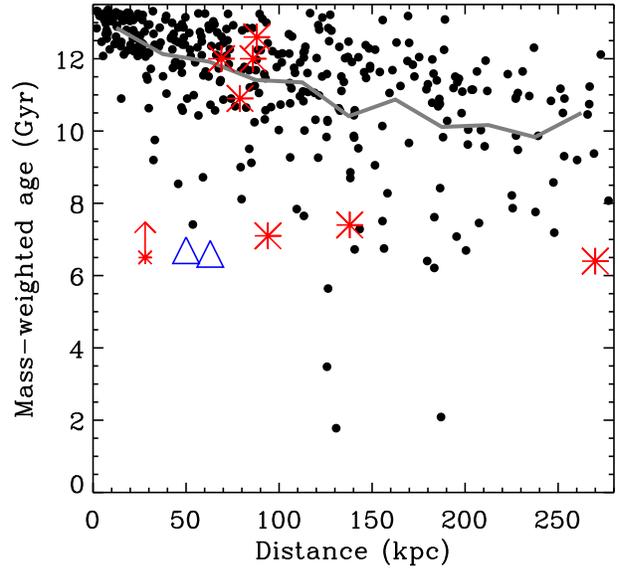}
\caption{Mass-weighted age as a function of their distance of the satellite to the main galaxy. The black circles show all Aquarius satellites with M$_{\rm{V}}<-5$ and the grey line represents the mean in this sample. Red asterisks represent mass-weighted ages for the Milky Way satellites from \citet{orba08}. Note that these are derived from HST observations and, because of the limited field of view and population gradients known in dwarf galaxies, might be biased towards lower mass-weighted ages. The two blue triangles are the Magellanic Clouds. The smaller red asterisk representing a lower limit is the Sagittarius galaxy, which has been severely stripped and therefore will have lost preferentially part of its older population. \label{oldyoung_rad}}
\end{figure}

Figure \ref{oldyoung_rad} shows that there is a clear trend in mass-weighted age with distance from the host. Systems dominated by intermediate populations are found preferentially in the outskirts of the host halo while older systems are found closer in. We see a similar trend in the Milky Way satellites, where the galaxies dominated by intermediate populations are located at greater distances than $\sim$ 100 kpc as indicated in Table \ref{tab:MWsats} (with the exception of the Magellanic Clouds and the heavily disrupted Sagittarius galaxy). Note that the mass-weighted ages for the Milky Way satellites plotted here are derived from HST observations, with a relative small field of view. These mass-weighted ages will therefore be biased towards the more concentrated younger population if a radial age gradient is present in the galaxy and metallicity and age gradients have been observed in many dwarf spheroidal systems \citep[e.g.,][]{harb01, tols04, batt06, bern08, debo12} as well as along the Sagittarius stream \citep[e.g.]{bell06}. 

\begin{table}
\begin{tabular}{lcccccl}
\hline
Name & M$_{\rm{V}}$ & \multicolumn{3}{c}{Detected Pops} & Dom. Pop & ref\\
& & old & im & yng & & \\
\hline
\hline
LMC & -18.1 & yes & yes & yes & old & [1]\\
SMC & -16.2 & yes & yes & yes & im & [2]\\
\hline
Sagittarius & -13.4 & yes & yes & no & ? & [3][4]\\
Fornax & -13.2 & yes & yes & no & im & [5][6]\\
\hline
Leo I & -11.9 & yes & yes & no & im & [3][7][8]\\
Sculptor & -11.1 & yes & yes & no & old & [3][9]\\
Leo II & -9.6 & yes & yes & no & ? & [3][7][10]\\
Sextans & -9.5 & yes & yes & no & old & [11][12][13]\\
Carina & -9.3 & yes & yes & no & im & [3][7][14]\\
Ursa Minor & -8.9 & yes & yes & no & old & [3][7][15]\\
Draco & -8.8 & yes & yes & no & old & [3][16]\\
CVn I & -8.6 & yes & yes & no & old & [17]\\
\hline
Hercules & -6.6 & ? & ? & ? & ? & \\
Bootes & -6.3 &  yes & yes & no & old & [17]\\
UmA I & -5.5 & ? & ? & ?& ? & \\
Leo IV & -5.0 & ? & ? & ?& ? & \\
\hline
CVn II & -4.9 & ? & ? & ? & ? & \\
Leo V &  -4.3 & ? & ? & ?& ? & \\
UmA II & -4.2 & yes & yes & no & old & [17]\\
Com Ber & -4.1 & ? & ? & ?& ? & \\
Boo II & -2.7 & ? & ? & ?& ? &\\
Willman 1 & -2.7 & ? & ? & ?& ? & \\
Segue 1 & -1.5 & ? & ? & ?& ? & \\
\end{tabular}
\caption{Populations in Milky Way satellite galaxies. References are: [1] \citet{harr09} (but note that the intermediate and old stars are almost contributing equally in the LMC SFH), [2] \citet{harr04}, [3] \citet{dolp02}, [4] \citet{bell99}, [5] \citet{cole08}, [6] \citet{gall05}, [7] \citet{hern00}, [8] \citet{gall99}, [9] \citep{debo12}, [10] \citet{gull08}, [11] \citet{mate92}, [12] \citet{bell01}, [13] \citet{lee03}, [14] \citet{hurl98}, [15] \citet{carr02}, [16] \citet{apar01}, [17] \citet{dejo08}.\label{tab:MWsats}}
\end{table}

\subsection{The physics shaping the modeled star formation histories}\label{sec:physSFH}

\begin{figure*}
\includegraphics[width=\linewidth]{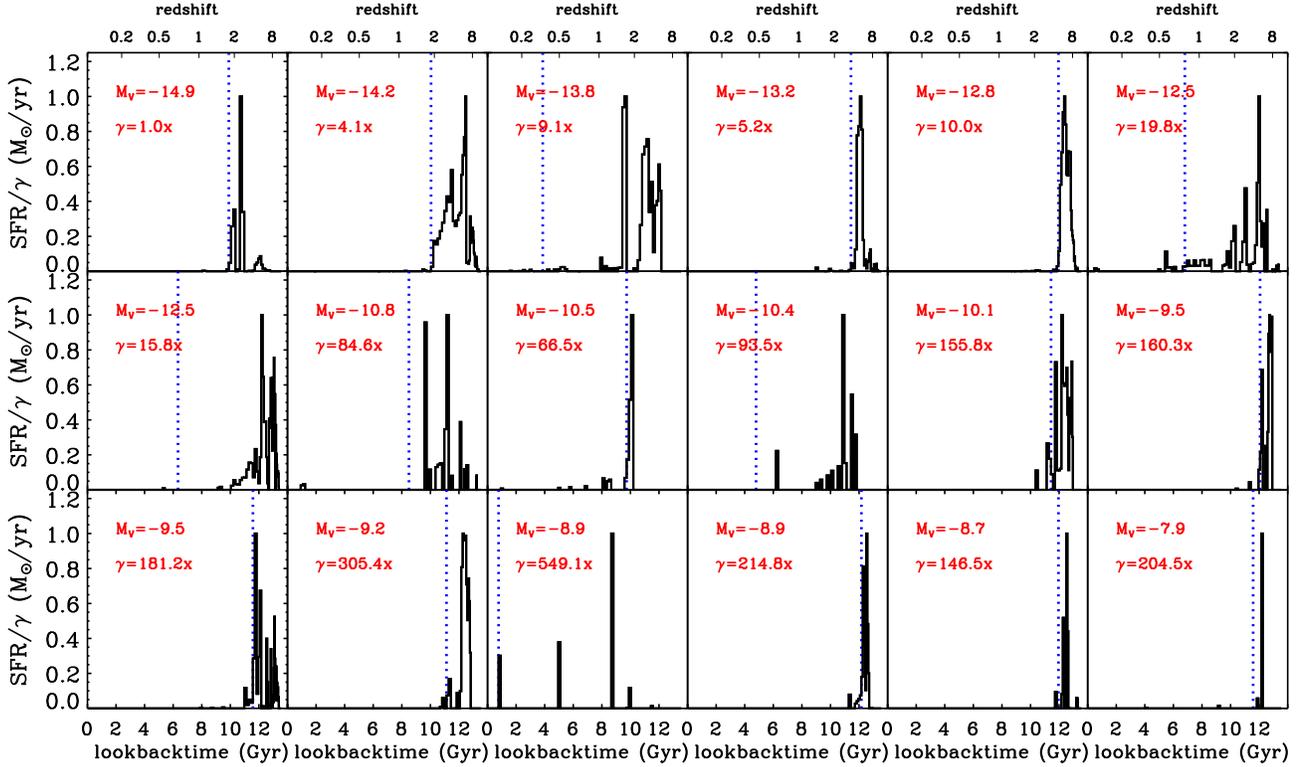}
\caption{The relative SFRs as a function of lookback time for the most luminous satellite galaxies in Aquarius B (black solid lines). We have indicated the infall time (blue vertical dotted line). In the top left corner of each panel, the absolute V-band magnitude of that particular galaxy is given and the scaling factor by which the star formation rates have been multiplied to obtain the normalized values.  \label{lookback}}
\end{figure*}

 Figure \ref{lookback} illustrates the variety of star formation histories found for the different satellites in Aquarius halo B. The 18 most luminous satellites are shown down to a magnitude of M$_{\rm{V}}$=-7.9. The star formation rates have been normalized in this figure to the highest peaks, but the absolute values range approximately from 0.2 M$_{\odot}$/yr in the top row to 0.005 M$_{\odot}$/yr in the bottom row. As can be seen from the figure, most stars in our satellites are made after reionization (at $z=11.5$ in our models), but before the time that the galaxy fell into the main halo (and thus became a satellite) as indicated by the vertical dashed blue lines. Taking all the stars in satellites brighter than M$_{\rm{V}}=-5$ and within 280 kpc in Aquarius haloes A-F, we find that 99.7 per cent of the stars formed after reionization, and 96.8 per cent before the satellite fell into the halo. The small satellites make a larger percentage of their stars before reionization. This is because they have just a few small bursts of star formation each of which contributes a significant fraction of the stars. In the larger galaxies the star formation rate is much higher and it extends over longer periods so that the initial episode of star formation (i.e. before reionization) is not important in a relative sense. On the other hand the quenching of star formation after infall ensures that a relatively small amount of stars will be made afterwards. 

Infall onto a host system in our model quenches star formation, because we
assume the hot halo of a galaxy is stripped as soon as the galaxy becomes a
satellite and the hot component is added to that of the host. Some models follow ram pressure stripping processes more
gradually \citep[e.g.][]{font08, wein10,guo11, nich11}. \citet{font08} and \citet{guo11} show that their
implementation does not make a significant difference for small systems as considered here. Some fully hydrodynamical simulations have also shown
that the remaining gas in the satellite galaxies is (almost) completely
stripped once a galaxy falls into the main halo \citep[e.g.][]{okam09,okam10}
and that the cooling of gas onto satellites seems to be small and important
only for massive systems \citep{saro10}. In our model, a satellite can still make some stars
after infall, but only as far as its cold gas reservoir allows it. We find that the amount of cold gas
still present in the modeled satellite galaxies, is in almost all cases larger than is observed within
satellite galaxies in the Milky Way halo \citep[e.g.][]{grce09}. However,
because this gas is below the star formation threshold, it does not form any
stars. It is quite possible that ram-pressure stripping of cold gas might have removed some of
this remaining gas from the observed satellite galaxies. Additionally, a more realistic model of star formation, not based on a global density threshold, but following the local density of the gas, might also alter the amount of gas left in these galaxies. The fact that all the classical dwarf spheroidals have a detected intermediate population, whereas this population is not always present in the models (see top left panel of Figure \ref{oldyoung}) hints that our model might shut off star formation too soon after infall.

A change in the value of the threshold density for star formation, equivalent to a more
  stochastic implementation to reflect molecular cloud physics, will not lead
to continuous star formation histories for most satellites, since the density of
cold gas is too far below the threshold most of the time. However, there are
some cases where the star formation threshold is barely not met, and a small
change in the value of the threshold, could significantly change the star
formation history. The most bursty star formation histories are found in
the galaxies which are constantly close, but mostly below, the star formation
threshold. 

\section{Milky Way dwarf analogs}\label{sec:findSclFnxCar}
\subsection{Star formation histories}
\begin{figure}
\includegraphics[width=\linewidth]{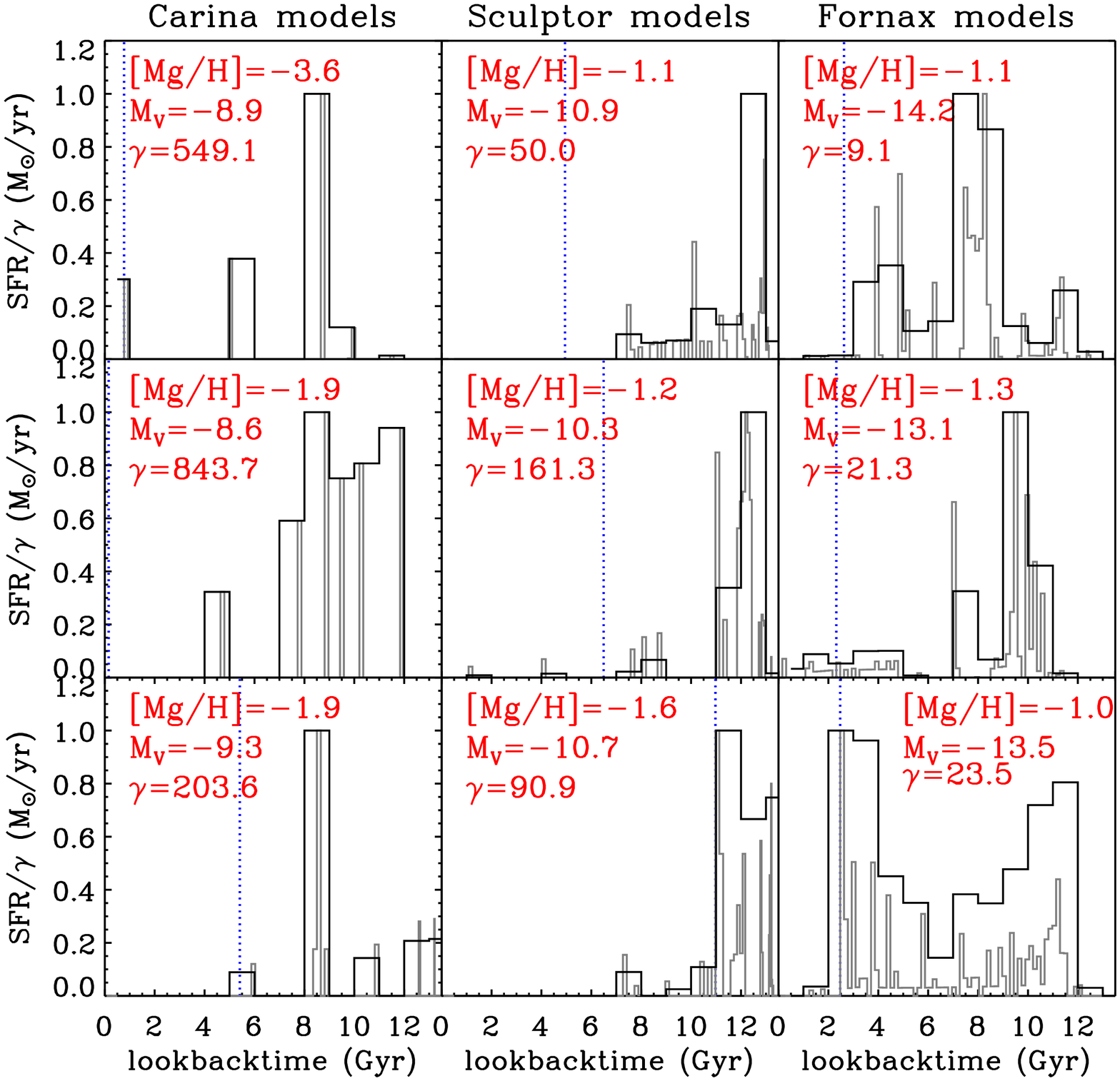}
\includegraphics[width=\linewidth]{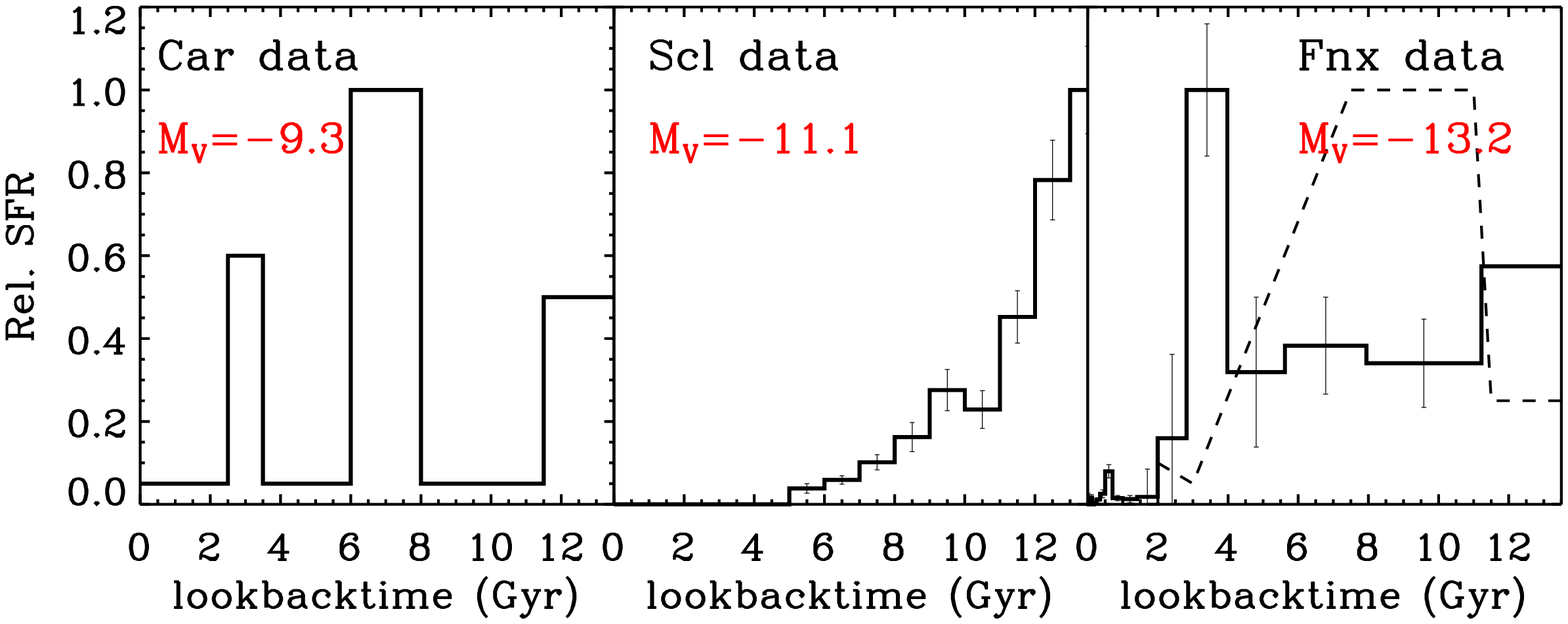}
\caption{Star formation histories for models which resemble the three classical dwarf galaxies Carina, Sculptor and Fornax (top panels). The star formation histories are shown both in the original binning of the simulation (grey lines) and rebinned to bins of 1 Gyr, to provide an easier comparison to the observations. In both cases the star formation histories are normalized to their peak value. Overwritten within the panels are the metallicity, luminosity and the scaling factor by which the model star formation rates have been multiplied to get the normalized values shown here. Division by this scaling factor thus returns the model star formation rate values, for the non-rebinned model (grey lines) in M$_{\odot}$/yr. We have also indicated the infall time of the satellite as a blue vertical dotted line. In the bottom panels we show observational star formation histories for the same galaxies (see text for references). \label{findscl}}
\end{figure}

In Figure \ref{findscl} we show some model satellite galaxies that are comparable in either brightness, average metallicity and star formation histories, or several of these properties, to the Carina, Sculptor and Fornax dwarf spheroidals. For each galaxy we show three examples to illustrate the scatter in these properties. 

In the bottom panels of Figure \ref{findscl}, we plot the observed star formation history for each dwarf spheroidal. Even though it is possible to observe all three galaxies down to their main sequence turnoff magnitudes, large differences are found in derived star formation rates due to incomplete spatial coverage or the set of model isochrones used \citep[see for example][]{gall05}. We have chosen to show here the star formation histories as obtained by \citet{cole08} for Fornax and by \citet{debo12} for Sculptor, because these are the most recent analyses, derived from photometry which covers (nearly) the entire galaxy. For Fornax, we also show, with a dashed line, the star formation as compiled by \citet{greb98}. The star formation history of Carina matches the one derived by \citet{hurl98}, with its characteristic three peaks. Other studies claim slightly different star formation histories \citep{dolp02,hern00}. Of course, one needs to bear in mind that the error bars on age are quite large for the observations (generally in the order of Gyrs for the regime we are most interested in), so the studies are not necessarily inconsistent. 

For Carina, we show several candidates selected on luminosity ($-10.5<$M$_{\rm{V}}<-8.5$) and their distinct bursty star formation with a majority of the stars formed at intermediate age. Amongst all our Aquarius simulations, we only find six candidates that have a dominant intermediate population in the corresponding luminosity bin. In Figure \ref{findscl} we show three candidates which have a very bursty star formation history. It is interesting that also bursty Carina-like galaxies are produced. In our models Carina's very bursty star formation history
occurs before it becomes a satellite and is the result of an interplay between gas density and star formation threshold, and is clearly not due to tidal interactions with the host \citep[such as suggested by][]{pase11}. All of these Carina-analogs fall into the main halo quite late. As discussed before in Section \ref{sec:physSFH}, most of these model galaxies find themselves close to the threshold of cold gas needed to make stars, and are only above the threshold occasionally. This explains the bursty nature of these star formation histories in our model. We have explicitly checked their merging history, but found no significant merging events causing or preceding any of the bursts. The Carina model shown in the bottom panel of Figure \ref{findscl}, depicts a star formation history that matches best the observed general properties of Carina, i.e. also matches the metallicity. Also this galaxy has a strong intermediate age burst of star formation, but a lower peak at a young age.

For Sculptor we found many analogs for the star formation history, since a great majority of our model satellite galaxies are dominated by an old population. However, most candidates of comparable luminosity have a slighter higher metallicity than Sculptor. Three typical examples of comparable luminosity, and also average metallicity (the largest offset is 0.5 dex), are shown in the middle column of Figure \ref{findscl}. 

The Fornax candidate model shown on the bottom panel of the right column of Figure \ref{findscl} matches very well the observed star formation history from \citet{cole08} and the general properties of the galaxy. On the other hand the models shown in the top and middle panels have star formation histories which match better older observations of Fornax in which the peak of star formation occurs at an older age, such as that compiled by \citet{greb98}. All examples shown here were chosen to have a comparable luminosity and average metallicity (the largest offset in metallicity is 0.2 dex).

\subsection{Metallicity distributions}

\begin{figure}
\includegraphics[width=\linewidth]{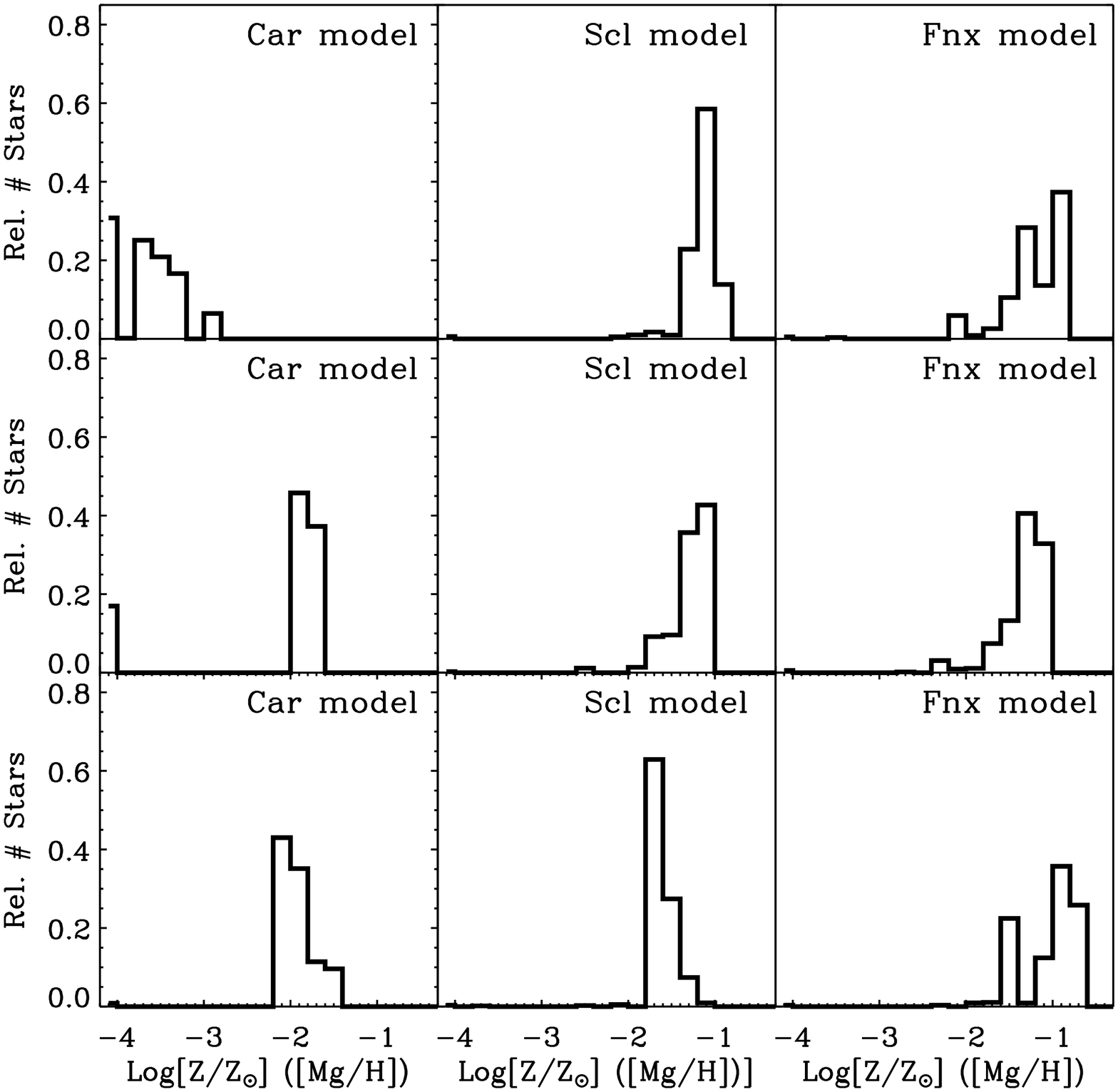}
\includegraphics[width=\linewidth]{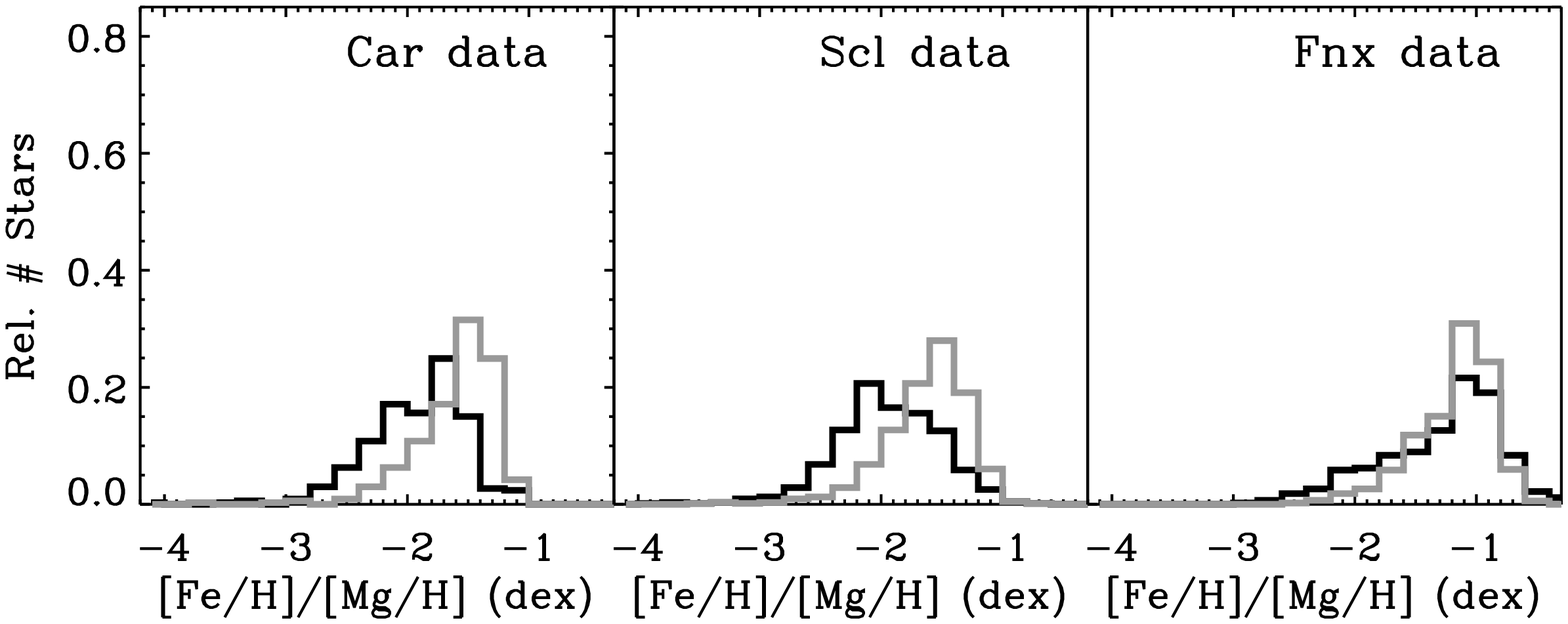}
\caption{Metallicity distributions for the models displayed in Figure \ref{findscl} which resemble the three classical dwarf galaxies: Carina, Sculptor and Fornax (top three panels). In the bottom panels we show observational metallicity distributions for the same galaxies (see text for references) in [Fe/H] from CaT samples (black solid lines), or corrected for a global relation to [Mg/H] values (grey solid line). \label{findscl_metdistr}}
\end{figure}

Figure \ref{findscl_metdistr} shows the metallicity distributions for the same set of model galaxies chosen to be relatively close analogs of the Milky Way dwarf spheroidals Carina, Sculptor and Fornax shown in Figure \ref{findscl}. The bottom panels show the observed metallicity distributions for these dwarf spheroidals as taken from the DART data sets. The black histograms correspond to the observed [Fe/H] distribution as derived from the Ca II triplet calibration of \citet{star10}. The grey histograms give the metallicity distributions corrected for [Mg/H] using the simplified relation of [Mg/Fe] discussed in Section \ref{sec:lum-met}. We regard this as a more direct comparison to our model, which assumed instantaneous recycling.

In general, the metallicity distribution functions of the model dwarfs are narrower, although a typical error on the observed metallicities is 0.2 dex, which will broaden the distribution. There is also a lack of extremely metal-poor stars in most metallicity distributions of our models. In the case of the Carina model shown in the top left panel however, too many very metal-poor stars are produced and the global metallicity is too low. This also is probably driven by our crude modelling of the first generations of stars, as discussed in Section \ref{sec:sfh}. Since the first peak of star formation is very small, it will only enrich the galaxy by a small amount leading to the formation of very low-metallicity stars in the second burst.

Overall, Figure \ref{findscl_metdistr} shows that the form of the distribution can vary significantly, depending on the exact star formation history of the system.

\section{Discussion}\label{sec:disc}

\subsection{Comparison to the {\sc galform} code} 

The Durham semi-analytic model of galaxy formation, {\sc galform} \citep[e.g.][]{cole00,bowe06}, has also been run on the same six Aquarius haloes by \citet{coop10} and by \citet{font11}. The latter implemented a novel and detailed treatment of both local and global reionization and compared the luminosity functions of their model and the satellite ejection model of \citet{li10}. This is identical to the one presented here in the bottom panel of Figure \ref{lumfuncABCDEF}, since in the comparison no orphan galaxies are considered and our stellar stripping mechanism does not significantly alter its shape. \citet{font11} concluded that both models are quite similar in methodology as well as results, including our treatment of reionization which is much less sophisticated as theirs. However, we note that a detailed treatment of local reionization could possibly affect also other observables such as the radial distribution of satellites, as shown by \citet{ocvi11}.

Our model and those resulting from {\sc galform} show some interesting differences however. For example the total stellar mass in the main galaxies are generally lower in the {\sc galform} models, ranging from $7 \times 10^{9}$ M$_{\odot}$ to $1 \times 10^{11}$ M$_{\odot}$ (compare Table 1 from \citet{font11} and Figure \ref{mainhalos} in this work). The number of satellites around the main halo are very similar, although there clearly are fewer very luminous satellites (brighter than M$_{\rm{V}}=-15$) present in the {\sc galform} models. This seems to be due to their slightly stronger feedback in this regime. In general the {\sc galform} satellite luminosity function is slightly steeper, but consistent with \citet{kopo08} and our results, in particular for luminosities fainter than M$_{\rm{V}}=-15$. Their luminosity-metallicity relation flattens out significantly at the fainter end (see their Figure 5, right panel), which is an interesting result directly related to their saturated feedback scheme, which is clearly distinct from the feedback scheme used in this work. Unfortunately, the errors on the available data are too large to test this prediction. 

\subsection{A stellar mass-halo mass relation: comparison to abundance matching and hydrodynamical simulations} 

An interesting issue in the context of the $\Lambda$CDM paradigm is which galaxies reside in which haloes. Several groups have quantified the relationship between stellar mass and halo mass using N-body simulations combined with (semi-analytical) models or observations \citep[e.g.][]{frenk88,nava00,deke03,yang03,wang06,shan06,yang08,conr09,most10,guo10}. This comparison is made under the assumption that the dark matter halo mass and the stellar mass of a central galaxy follow a monotonic relation, taking into account some amount of scatter. For example \citet{guo10} combined the abundance of galaxies with stellar masses in the range $10^{8}\rm{M}_{\odot}<M_{*}<10^{12}\rm{M}_{\odot}$ from SDSS \citep{liwh09}, and the dark matter halo abundances from the Millennium Simulations \citep{spri05, boyl09}. The relation derived from this analysis predicts a dark matter mass for a Milky Way-type galaxy of $\sim2\times10^{12}$M$_{\odot}$, which is a factor of two larger than the favoured result from our semi-analytic model, but within the current observational constraints. 

\begin{figure}
\includegraphics[width=\linewidth]{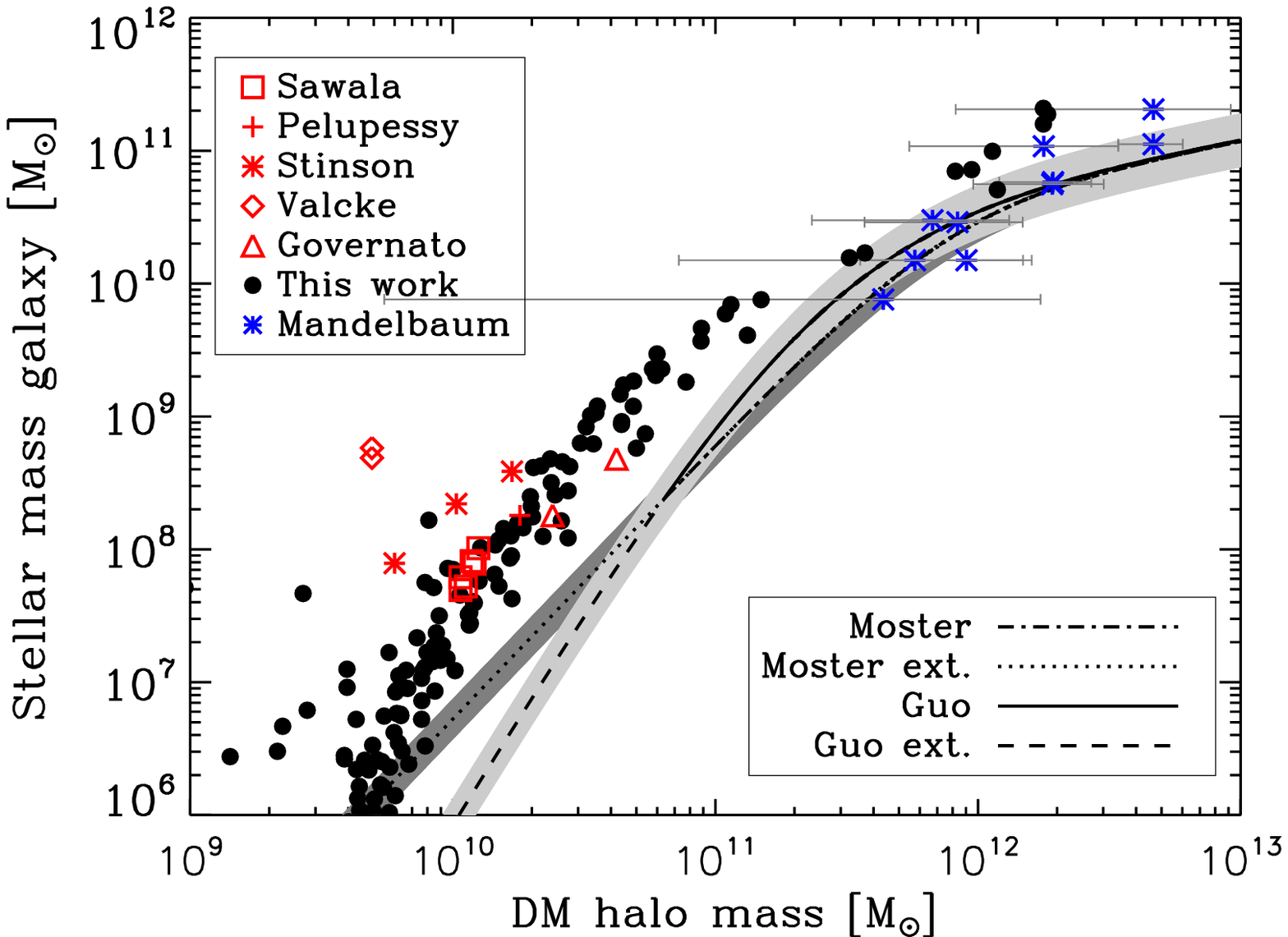}
\caption{The relation between stellar masses and dark matter halo masses for all central haloes within 2.5 Mpc from the main halo in the Aquarius simulations (black filled circles). The red symbols are results from hydrodynamical simulations, as compiled in \citet{sawa11b}, from \citet{pelu04,stin07,gove10,sawa11b}. The blue asterisks are galaxies from an SDSS sample for which stellar masses were derived from spectroscopy and dark matter virial halo masses from weak gravitational lensing by \citet{mand06}; the error bars give the 95 per cent confidence intervals. The black solid line represents the stellar mass-halo mass relation as derived by \citet{guo10}, constrained by SDSS DR7 data. The extrapolation into the low-mass regime is indicated by the black dashed line, whereas the light grey area shows the maximum dispersion, $\sigma_{\rm{logM_{*}}}=0.2$ dex. The stellar mass-halo mass relation derived by \citet{most10}, including a scatter of $\sigma_{\rm{logM_{*}}}=0.15$ dex (dark grey area), is shown as a black dashed-dotted line and its extrapolation into the low-mass regime is given by the dotted black line. \label{whichinwhich}}
\end{figure}

\citet{sawa11b} noticed that current hydrodynamical simulations of isolated dwarf galaxies produce an order of magnitude larger stellar masses than the expected relation from \citet{guo10} extrapolated to the lower halo mass end. In Figure \ref{whichinwhich} we show the relation from \citet{guo10} and \citet{most10} as well as the results predicted by our model for all the central galaxies in the Aquarius simulations\footnote{We have used a distance limit of 2.5 Mpc from the centre of the box to stay well within the high-resolution region of our simulations.} and the hydrodynamical simulations from \citet{pelu04,stin07,gove10,sawa11b}, as compiled by \citet[][see also their Figure 4.]{sawa11b}. 

We see a strong correlation between total stellar mass and dark matter halo mass \citep[taken as their virial mass at $z=0$, as in][]{guo10} for the central galaxies in our model, as shown in Figure \ref{whichinwhich}. However, the relation predicted from our model and the one predicted by \citet{guo10} and \citet{most10} are clearly offset: at a given luminosity our model galaxies reside in smaller dark matter haloes. The difference is greatest for the lower mass objects, for which the \citet{guo10} and \citet{most10} relations are extrapolations (i.e. these scales are poorly constrained by the data used by these authors; and are indicated by the dashed and dotted lines in Figure \ref{whichinwhich}), but also at the regime of the main central galaxies (our `Milky Way's) there is a significant difference (with the exception of the main galaxy in Aquarius E). Generally, our results are in agreement with \citet{mand06}, who used SDSS to derive stellar masses from spectroscopy and virial halo masses from weak gravitational lensing. Interestingly, while being offset from the results of \citet{guo11} and the relation from \citet{guo10} at the lower masses, our model predictions are also consistent with most hydrodynamical simulations shown.   

Both \citet{most10} and \citet{guo10} demonstrate that the relation for the most likely stellar mass within a given dark matter halo changes with different assumptions about scatter. The scatter within the Aquarius haloes is larger than the predictions from either \citet{guo10} and \citet{most10}, certainly at the lowest masses, raising the question of whether environmental stochastic effects (such as those associated to the time of formation, the mass at reionization, the cold gas density and the star formation threshold) may play a more important role for low mass systems. 

Recent work of \citet{vera12} and \citet{wang12b} show that the assumption of a lower mass for the Milky Way around $\sim8 \times 10^{11}$ will alleviate significantly the problem raised by \citet{boyl11} that satellite systems around Milky Way-sized haloes would be too dense to host the known Milky Way dwarf galaxy satellites. Using the model presented in this paper \citet{vera12} subsequently show that good agreement can be found for the relation between luminosity to total mass of the modeled satellites in comparison with observational constraints \citep[in disagreement with the V$_{\rm{max}}$-M$_{\rm{V}}$ dichotomy between modeled and observed satellites as shown in Figure 6. of][based on abundance matching]{boyl12}.

Using an extension of the \citet{delu07} model, \citet{guo11} obtain a good correspondence with the relation derived by \citet{guo10} for SDSS galaxies and consistency with galaxy luminosity functions over a large range in magnitudes in various bands and with the extension of the \citet{guo10} relation for lower mass galaxies. They found that in order to reach this good correspondance they needed to adopt a supernova feedback ejection efficiency which depends on the circular velocity of the underlying dark matter halo with an exponent $\beta_{\rm{1}}=-3.5$, whereas in our model $\beta_{\rm{1}}=-2$. A careful comparison of both models on larger cosmological scales is beyond the scope of this paper and also not feasible with the set of simulations used, but we plan to address this issue in future work.

\section{Conclusions}\label{sec:Aqconc}
In this work we have confirmed that a semi-analytic model of galaxy formation applied to a high resolution cosmological N-body simulation is able to
match observed relations on the scale of the Milky Way and its satellites
simultaneously. We have compared the results for the luminosity function,
luminosity-metallicity relation and radial and spatial distribution to observations of
dwarf galaxies around the Milky Way and investigated the age-density relation and star formation histories within the model. 

We find that the same Aquarius halo in which we found the closest Milky Way-like galaxy analogue, halo B, also provides us with the best matching satellite system in terms of its luminosity function. This galaxy does not resemble the Milky Way in every respect. For example the radial distribution  of satellites within this particular simulation is more centrally concentrated than the observed distribution of satellites around the Milky Way and it does not show the same degree of spatial flattening. It also does not host a galaxy with a luminosity comparable to the LMC or SMC. Of course, the Milky Way, although in many ways the best studied galaxy we know, is just one example and it might be unusual in several respects \citep[e.g.,][]{flyn06,guoquan11}. 

We find a clear relation in our models between the number of bright satellites and the host dark halo mass. With our current feedback and reionization prescriptions, our best Milky Way analogue has a dark halo close to 8 $\times 10^{11} \rm{M}_{\odot}$, in agreement with the lower estimates from observations \citep{batt05,smit07,xue08}, but a factor of two lower than the best estimates from \citet{liwh08,guo10}. Additionally, over the mass range probed by the Aquarius simulations, we find a different relation between dark matter halo mass and stellar mass of the central galaxies than that derived by \citet{guo10} and \citet{most10}, and reproduced by \citet{guo11}.  

Based on their star formation histories, we find model satellites analogous to
the Sculptor, Fornax and Carina dwarf spheroidals, although none of the model
galaxies provides a match of all observable properties. However, the
metallicity distributions for these galaxies are generally too narrow compared
to the observations and they lack an (extremely) metal-poor population. It is
unclear at the moment whether this can be completely ascribed to the lack of a
detailed prescription of the chemical evolution of different elements
and to the adoption of the instantaneous recycling approximation. This topic will be the subject of
further work. Also our model does not allow cooling via H$_{2}$ in haloes below
the Hydrogen atomic cooling limit, and hence does not provide a fully physical model for the
formation of the first stars. Another shortcoming of our current model is that
our modeled satellite galaxies have too much cold gas compared to
observations. This could perhaps be (partly) solved by including ram-pressure
stripping of the cold gas when an object becomes a satellite.

However, various predictions can be made from our model that are expected to be independent of these shortcomings. We predict the ratio between galaxies dominated by old- or intermediate populations of stars to be close to 1:2 beyond the Local Group. We also expect a large majority of the satellite galaxies that are dominated by intermediate age stellar populations, to have fallen into the main halo relatively late, including galaxies with a bursty star formation like Carina. Generally, a very small percentage of stars in a satellite is formed after its infall-time, due to stripping of the satellite's hot halo which prevents further gas from cooling and forming stars. However, some galaxies within the model have stopped star formation several Gyrs before their infall, due to internal processes, such as supernova feedback, which have expelled their gas. Our model thus predicts the last major star formation event within a (classical) satellite to rather provide a lower limit on the time elapsed since infall.

The brighter galaxies amongst the ultra-faint satellites (e.g $-8.5<$M$_{\rm{V}}<-5$), which could be well resolved in our model, have formed a larger percentage of their stars in a single burst, because feedback has a larger impact and prevents a continuous mode of star formation. These model ultra-faint galaxies are generally older than the more luminous counterparts, and hence contain a higher fraction of stars formed around, and even before, the epoch of reionization.

\section*{Acknowledgments}
We thank Eline Tolstoy, Stefania Salvadori and Andrew Cooper and the referee for very helpful suggestions that helped improve the paper. E.S., A.H. and C.V.-C. thank the Netherlands Foundation for Scietific Research (NWO) and the Netherlands Research School for Astronomy (NOVA) for financial support. E.S. also gratefully acknowledges the Canadian Institute for Advanced Research (CIfAR) Junior Academy and the Canadian Institute for Theoretical Astrophysics (CITA) National Fellowship for partial support, and the organisers of the KITP programme “First Galaxies and Faint Dwarfs: Clues to the Small Scale Structure of Cold Dark Matter” for a stimulating work environment where much progress on this article has been made. As such, this research was supported in part by the National Science Foundation under Grant No. NSF PHY11-25915. A.H. was supported by the ERC-StG Galactica 240271. G.D.L. acknowledges financial support from the European Research Council under the European Community's Seventh Framework Programma (FP7/2007-2013)/ERC grant agreement n. 202781. C.S.F. acknowledges a Royal Society Wolfson Research Merit award and is supported in part by ERC Advanced Investigator grant COSMIWAY. This work was also supported in part by an STFC rolling grant to the ICC. V.S. acknowledges support through SFB 881, `The Milky Way system', of the DFG.

\bibliography{ms}

\appendix

\section{Implementation of stellar stripping and tidal disruption}\label{sec:strip}

\subsection{Stellar stripping}\label{sec:stelstrip}
In our simulations, the dark matter subhalo of a satellite may be so heavily tidally stripped that also its stellar component should be affected. Here we implement an approach to ensure that no very extended satellite galaxies reside within much smaller, heavily stripped dark matter subhaloes. This approach is thus only used on satellites which still have a dark matter component. 

The half-mass radius of an exponential disc is related to its scale-length through $R^{*}_{\frac{1}{2}}= 1.67 \times R_{\rm{d}}$. At each time-step we now compare the half-mass radius of the stars (and cold gas), $R^{*}_{\frac{1}{2}}$, to the half-mass radius of the dark matter subhalo, $R^{\rm{DM}}_{\frac{1}{2}}$. The value of $R^{\rm{DM}}_{\frac{1}{2}}$ is measured by the {\sc subfind} algorithm \citep{spri01} directly using the dark matter particles belonging to the satellite. If through tidal stripping in the simulations $R^{\rm{DM}}_{\frac{1}{2}} < R^{*}_{\frac{1}{2}}$, we remove the stars (and corresponding cold gas) up to $R^{DM}_{\frac{1}{2}}$. This implies that after stripping the disc now has a mass given by
 \begin{equation}
 M_{\rm{d-new}}=2M_{\rm{d}}\bigl[1-\bigl(1+\frac{R^{\rm{DM}}_{\frac{1}{2}}}{R_{\rm{d}}}\bigr)e^{-R^{\rm{DM}}_{\frac{1}{2}}/R_{\rm{d}}}\bigr].
\label{disc}
\end{equation}
The updated exponential disc, consisting of the leftover cold gas, has a half-mass radius set by $R^{*}_{\frac{1}{2}-new} = R^{\rm{DM}}_{\frac{1}{2}}$ and the new scale radius for the disc is therefore $R_{\rm{d-new}} = R^{\rm{DM}}_{\frac{1}{2}}$/1.67.

The cold gas and stars stripped from the satellite, are transferred
respectively to the cold gas disc and spheroidal stellar component,
which includes the bulge and stellar halo, of the main galaxy.

Although the assumption that 50 per cent of the stars should at least be contained within the dark matter half-mass radius seems like a natural first guess, there are very little direct observational constraints on the exact ratio. However there is general agreement that the dark matter halo indeed has to be severely stripped before stars are affected. For instance, the scale length of the Milky Way stellar thin disc is $\sim 3$ kpc, whereas its dark matter half-mass radius is thought to be in the order of 100 kpc. For a more in dept discussion of tidal stripping in disc galaxies, and the influence of parameters as galaxy mass, orbit and disc inclination from numerical simulations we refer the reader to \citet{vill12}. 

\subsection{Tidal disruption}\label{sec:tiddisr}
In our tree files a dark matter subhalo is `lost' when it is stripped down to less than 20 particles \citep[corresponding to a dark matter mass of $\sim2.7\times10^{5}$ M$_{\odot}$ in the Aquarius A halo simulated at level 2][]{spri08a}. We refer to these galaxies as `orphan'. The question arises: when their dark matter subhalo disappears what should be their fate. 

We assume that the satellite galaxies residing at the centres of the fully (below the resolution limit) stripped subhaloes will merge with the central galaxy after a dynamical friction time scale, as explained by \citet{delu08}. However, for low mass galaxies this time scale will be longer than the Hubble time. In some previous work on semi-analytical models of Milky Way haloes \citep[e.g.][]{li09,li10,font11} these orphan satellites galaxies were not taken into account, since they were believed to get completely stripped. However this needs to be checked explicitly and in some cases we can expect the galaxies to survive the tidal forces.

Here we improve further the code, by comparing the average mass density of orphan satellites to that of their host system and determine whether they will survive. We do this \textit{only} for those orphan galaxies that would otherwise survive, so will not merge on a dynamical friction time-scale with the central galaxy before $z=0$. This means we treat the more and less massive orphan galaxies via different mechanisms. The more massive orphans are dragged in and merge with the central at the dynamical friction time-scale, the less massive might disrupt within the halo (or not, depending on their density). Differences between the two implementations appear firstly in the time-scale of merging (which is set at the dynamical friction time-scale for the massive orphans and just taken as the first snapshot at which they become orphans for the less massive galaxies) and secondly their impact on the host galaxy: only within the merging prescription black hole growth and the impact of major mergers are modeled (which can induce starbursts and morphological changes). Tests showed that if all orphans be treated such to disrupt within the halo, the final properties of the host galaxies can be affected. In most cases, the changes are small and mainly affect the bulge-to-disc ratio and the mass of the central black hole.

For the disruption mechanism we compare the density of these satellites, at the time they become a orphan galaxy, to the density of the host galaxy at pericenter. In the case that the density of the satellite is not sufficiently high compared to the environment, they are (assumed to be) completely disrupted. A similar tidal disruption mechanism was already followed in several semi-analytical codes \citep[e.g.][]{mona07,henr08,guo11}. The approach described here is in essence most comparable to the implementation of \citet{guo11}, but there are some significant differences, most importantly we use a Navarro, Frenk and White profile \citep[NFW;][]{nava96} for the host halo whereas \citet{guo11} assume an isothermal sphere.

To establish whether a orphan satellites will survive or not, we need to estimate its density at the pericenter of its orbit. This distance may be derived numerically from:
\begin{equation}
E_{\rm{tot}} = \frac{1}{2}\frac{L^2}{R_{\rm{peri}}^2} + \Phi_{\rm{NFW}}(R_{\rm{peri}}),
\label{peri}
\end{equation}
here the total energy, $E_{\rm{tot}}$, and angular momentum, $L$, are those from the last recorded time.

Subsequently the density of the satellite, $\langle\rho_{\rm{sat}}\rangle$ is compared to the average density of the host, $\langle\rho_{\rm{host}}\rangle$, at the pericenter distance:
\begin{equation}
\langle\rho_{host}\rangle = \frac{M_{halo}(R_{peri}) + M_{*} + M_{gas}}{R_{peri}^{3}}.
\label{host}
\end{equation}
 Since the satellite's dark matter content has fallen below the resolution limit of the N-body simulation it is not possible to evaluate its density precisely. Here we adopt two approaches to bracket the true density of the satellite.
\begin{itemize}
\item{To obtain an upper limit to the satellite's density ($\langle\rho_{sat}\rangle$), we measure the average dark matter density within $R^{\rm{DM}}_{\frac{1}{2}}$ in the last snapshot the satellite was still detected within the simulation. To compute the satellite's average density, we also take into account a fraction of the stellar and cold gas mass of the satellite galaxy. This fraction of baryonic mass within $R^{\rm{DM}}_{\frac{1}{2}}$ is determined as:
\begin{equation}
 f=\frac{M_{\rm{d}}(R^{\rm{DM}}_{\frac{1}{2}})}{M_{\rm{d}}}=\bigl[1-\bigl(1+\frac{R^{\rm{DM}}_{\frac{1}{2}}}{R_{\rm{d}}}\bigr)e^{-R^{\rm{DM}}_{\frac{1}{2}}/R_{\rm{d}}}\bigr],
\label{disc2}
\end{equation}
\begin{equation}
\langle\rho_{\rm{sat}}\rangle = \frac{M_{\frac{1}{2},\rm{sat}}^{\rm{DM}} + fM_{\rm{sat}}^{*} + fM_{\rm{sat}}^{\rm{gas}}}{(R_{\frac{1}{2},\rm{sat}}^{\rm{DM}})^{3}}.
\label{disruptdm}
\end{equation}
}
\item{To obtain a lower limit to the satellite's density, we assume all dark matter is stripped and we take as the average satellite density that given by the full stellar and cold gas mass of the galaxy within five disc scale lengths (within which 95 per cent of the stars should be contained). In this case:
\begin{equation}
\langle\rho_{\rm{sat}}\rangle = \frac{M_{\rm{sat}}^{*,\rm{gas}}}{(5R_{\rm{d,sat}})^{3}}.
\label{disrupt}
\end{equation}
}
\end{itemize}
We consider the satellite galaxy to be disrupted when $\langle\rho_{\rm{sat}}\rangle < \langle\rho_{\rm{host}}\rangle$, where $\langle\rho_{\rm{sat}}\rangle$ is given by either Eq. (\ref{disruptdm}) or (\ref{disrupt}). The cold gas component of the disrupted satellites is added to that of the main galaxy. We tested that the addition of this gas to either the cold or the hot gas component of the main galaxy does not significantly affect the properties of the main galaxies, since it might not be completely physical to assume that all the cold gas from the satellite flows into the cold gas reservoir of the host. The stars from the disrupted satellites are added to the spheroid component (which includes the stellar halo of the main galaxy).

\subsection{The effects of the stripping and disruption prescriptions}\label{sec:stripeff}

\begin{figure*}
\includegraphics[width=0.36\linewidth]{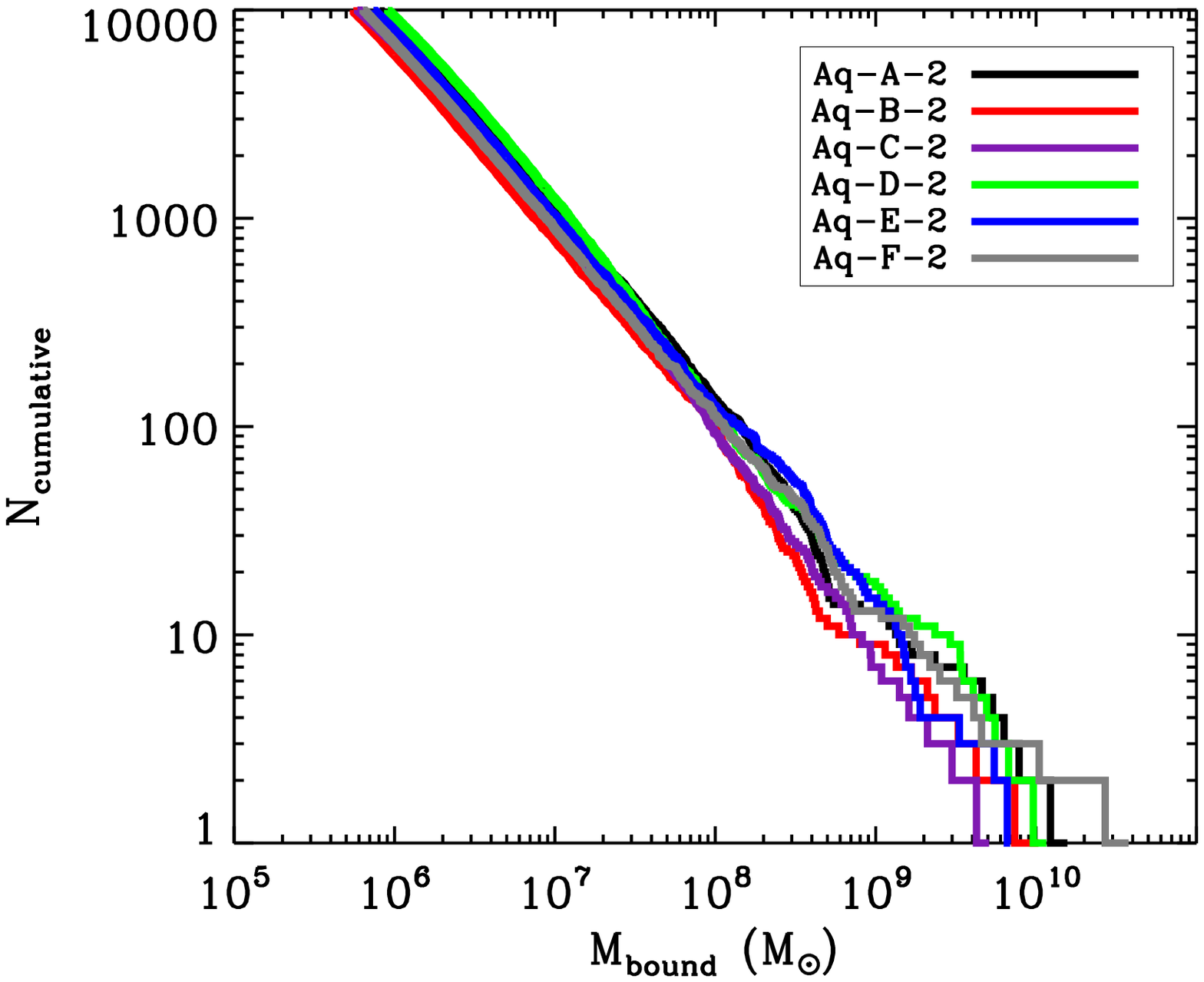}\includegraphics[width=0.64\linewidth]{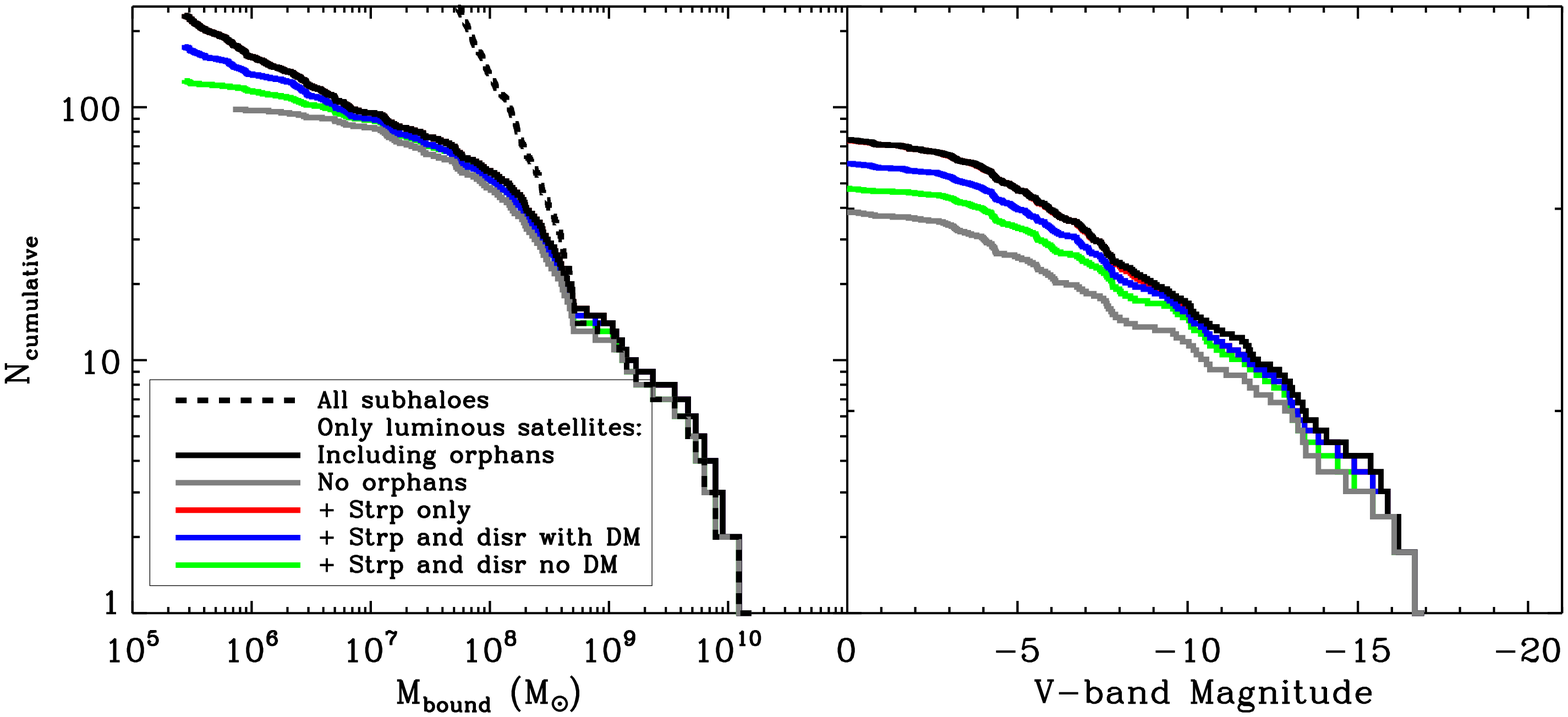}
\caption{Leftmost panel: Mass functions for all systems of subhaloes within 280 kpc of all six Aquarius haloes. Right panels: Mass (left) and luminosity (right) functions for all luminous satellite galaxies within 280 kpc of the main galaxy in Aq-A-2. Different models are used: the default model \citep[`ejection' model of][black solid line]{li10}, only the galaxies with a dark matter subhalo within the default model (grey solid line), default model now including stellar stripping (red solid line), and default model including both stripping and tidal disruption with or without dark matter (providing an upper and lower limit to the luminosity function, blue and green solid lines respectively). Additionally, the mass function of the system of all subhaloes is overplotted as a dashed black line in the left panel. The effect of stellar stripping alone, shown here as the difference between the black and red solid line, is not visible in the mass function (as one would expect) and has only a small effect on the luminosity function. Notice that the masses shown are the present-day masses obtained by adding all bound particles of a subhalo. This value can be affected by stripping processes and therefore be significantly reduced from the mass at infall. For the orphan galaxies the mass is used from the last snapshot in which they were found. \label{lumfuncstrpdisr}}
\end{figure*} 

In Figure \ref{lumfuncstrpdisr} we show the mass functions of present-day masses for the systems of subhaloes, and the mass- and luminosity functions of luminous satellites in Aq-A-2 using different modeling prescriptions. From the middle panel of the figure, where both the total subhalo and luminous subhalo mass functions are overplotted for Aq-A-2, it is clear that all massive halos at present day (M$_{\rm{bound}} > \sim$10$^{9}$M$_{\odot}$) will have formed stars, but that at the lower mass end subhalos with similar present-day masses (which can be affected also by stripping processes) might or might not have a luminous component.

Figure \ref{lumfuncstrpdisr} also shows the effect of the stellar stripping and tidal disruption prescriptions, introduced in Section \ref{sec:stelstrip} and \ref{sec:tiddisr}, on both the mass function and the luminosity function in Aq-A-2. Overall, the stripping mechanism implemented in the model has a small effect on the total luminosity function, only affecting a small percentage of the bigger satellite galaxies (M$_{\rm{V}} \sim -10$). This is still true if we relax our assumption on the percentage of stars to be contained within the dark matter half-mass radius, and change this from 50 per cent to 20 or 80 per cent. In Aq-A-2, just 3 present-day galaxies with M$_{\rm{V}} < -5$ are affected by the stripping mechanism. Nonetheless, the stellar stripping criterion is an important addition for the study of individual galaxies since it prevents the unphysical presence of too massive galaxies within very heavily stripped dark matter subhaloes in the model. It also strips down galaxies which will become orphans eventually, making the comparison of their densities to that of the halo more realistic. 

The implementation of tidal disruption affects the orphan satellites, but not all of them, as shown in Figure \ref{lumfuncstrpdisr}. 

\section{The effects of numerical and time resolution}\label{sec:numres}

\begin{figure}
\includegraphics[width=\linewidth]{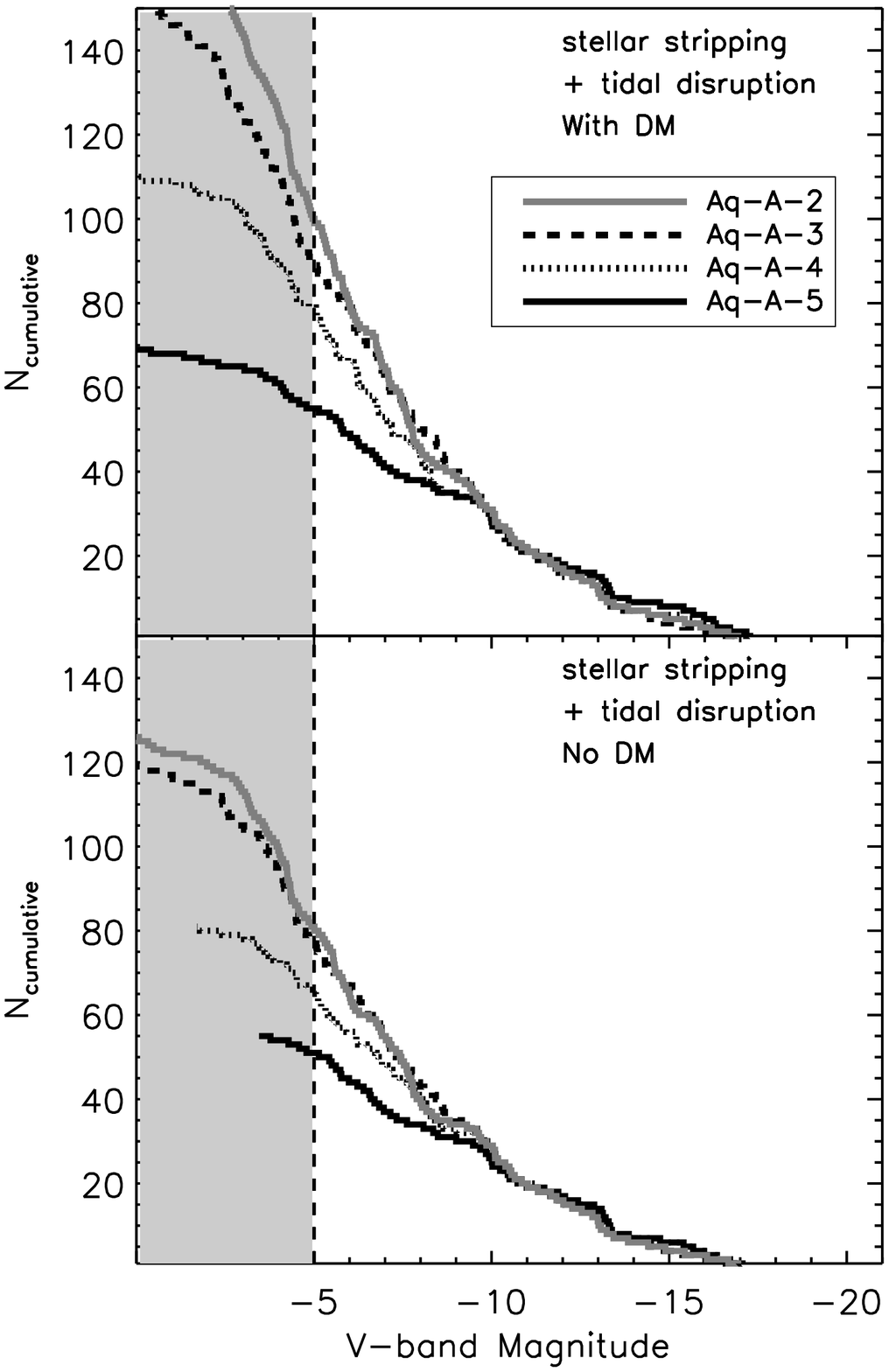}
\caption{Luminosity function for Aquarius A levels 2-5 for the two ways of estimating the average density of a satellite for tidal disruption (see Section \ref{sec:stelstrip}). In the top panel the density estimate includes the contribution of the dark matter, while in the bottom panel it does not. The dashed vertical line and grey area in the bottom panels indicate the luminosity of satellite galaxies at which Aq-A-2 and Aq-A-3 start to deviate and thus the numerical resolution be considered to affect the results. \label{lumfuncstrp}}
\end{figure} 

\subsection{Numerical resolution}
The Aquarius Project haloes have been run at different numerical resolution levels and these runs show remarkable good convergence on the properties (mass, position and kinematics) of the simulated haloes \citep{spri08a}. This enables us to explore the effects of numerical resolution on our models very directly. Figure \ref{lumfuncstrp} shows the effect of numerical resolution for both tidal disruption approaches on the luminosity functions. The luminosity functions of Aq-A-2 and Aq-A-3 start to diverge significantly for M$_{\rm{V}}>-5$. This justifies the choice used throughout the article to use this magnitude limit in our comparisons to the Milky Way satellites, which includes many of the (brighter) ultra-faint satellites. Lower level resolution merger trees are only available for halo A, we can therefore not perform similar convergence tests for the other simulations.

We also choose to use the disruption prescription without dark matter (i.e., based on the lower limit estimate of the density of the satellite) in the rest of this work. Either choice would be an approximation, due to the intrinsic uncertainties in the modelling of orphan galaxies. However Figure \ref{lumfuncstrpdisr} shows that the tidal disruption without dark matter shows a convergence between Aq-A-2 and Aq-A-3 down to fainter magnitudes. At the level of M$_{\rm{V}}=-5$ the lower limit approach predicts the disruption of 20 additional galaxies (see Figure \ref{lumfuncstrpdisr}), while in Aq-B the difference is only 7 systems, since this halo has a much smaller population of orphan galaxies.

\section{Resolution of timesteps}
Additionally, we test the dependence of the star formation histories obtained in the model for their dependence on the resolution of timesteps taken. We use Aq-A-4 for this purpose, which has a very large number of snapshots and can thus be rebinned into different time resolutions. We find that the less luminous galaxies, which often have the more bursty star formation histories, are the most vulnerable to a change in the number of time-steps taken in the simulation. By changing the time resolution in the simulation by a factor of two and rebinning the final output to the lowest time resolution, typically $\sim$30 per cent of the stars in a dwarf galaxy will be formed in a different snapshot. For the galaxies with M$_{\rm{V}}<-8.5$ only, this shrinks to 25 per cent. In all cases, this has hardly any effect on the characterization of the star formation history in old, intermediate, or young bins as used in this paper. Only typically $\sim$1 per cent of stars ever will change between these bins in the tests. We have also checked the time resolution within the semi-analytical model itself and found that the resolution used is sufficient; i.e. by increasing the number of time steps by a factor 2, changes to the results as used in the paper are negligible.

\end{document}